\documentclass[useAMS,usenatbib,usegraphicx]{mn2e}

\usepackage{amssymb}
\usepackage{float}
\usepackage[font=footnotesize,caption=false]{subfig}
\usepackage{booktabs}
\usepackage{ulem}
\usepackage[usenames,dvipsnames]{color}

\newcommand{\nar}{New A Rev.}
\newcommand{\mnras}{MNRAS}
\newcommand{\nat}{Nat}
\newcommand{\aj}{AJ}
\newcommand{\apj}{ApJ}
\newcommand{\apjl}{ApJL}
\newcommand{\apjs}{ApJS}

\newcommand{\aap}{A\&A}

\newcommand{\pasp}{PASP}
\newcommand{\araa}{ARA\&A}

\def\Msun{\hbox{M$_{\odot}$}}
\def\sfrunits{\Msun\ yr$^{-1}$}
\def\halpha{H$\alpha$  }
\def\halphans{H$\alpha$}
\def\hbeta{H$\beta$  }
\def\smean{$\sigma_{\mathrm{mean}}$}
\def\vshear{$v_{\mathrm{shear}}$  }
\def\NII{[N\hspace{.03cm}II]}

\def\OIII{[O\hspace{.03cm}III]}
\def\HII{H\hspace{.03cm}{\small II}~}
\def\stromgren{Str\"{o}mgren}

\title[Scaling Relations of Star-Forming Regions]{Scaling Relations of Star-Forming Regions: from kpc-size clumps to H{\LARGE II} regions}
\author[E. Wisnioski et al.]{Emily Wisnioski$^{1}$\thanks{E-mail: ewisnios@astro.swin.edu.au},
Karl Glazebrook$^{1}$, 
Chris Blake$^{1}$, 
Gregory B.\ Poole$^{1}$,\newauthor
Andrew W. Green$^{1,2}$, 
Ted Wyder$^{3}$, 
Chris Martin$^{3}$\\
$^{1}$Centre for Astrophysics and Supercomputing, Swinburne University of Technology, P.O. Box 218, Hawthorn, VIC 3122, Australia\\
$^{2}$Australian Astronomical Observatory, P.O. Box 296, Epping, NSW 2121, Australia\\
$^{3}$California Institute of Technology, MC 405-47, 1200 East California Boulevard, Pasadena, CA 91125, United States}

\begin{document}

\date{Accepted 2012 February 29. Received 2012 February 13; in original form 2011 December 13.}

\pagerange{\pageref{firstpage}--\pageref{lastpage}} \pubyear{2011}

\maketitle

\label{firstpage}

\begin{abstract}
We present the properties of 8 star-forming regions, or `clumps,' in 3 galaxies at $z\sim1.3$ from the WiggleZ Dark Energy Survey, which are resolved with the OSIRIS integral field spectrograph. Within turbulent discs, $\sigma\sim90$ km s$^{-1}$, clumps are measured with average sizes of $1.5$ kpc and average Jeans masses of $4.2\times10^9$ \Msun,~in total accounting for $20-30$\% of the stellar mass of the discs. These findings lend observational support to models that predict larger clumps will form as a result of higher disc velocity dispersions driven-up by cosmological gas accretion. As a consequence of the changes in global environment, it may be predicted that star-forming regions at high redshift should not resemble star-forming regions locally. Yet despite the increased sizes and dispersions, clumps and \HII regions are found to follow tight scaling relations over the range $z=0-2$ for \halpha size, velocity dispersion, luminosity, and mass when comparing $>2000$ \HII regions locally and 30 clumps at $z>1$ ($\sigma \propto r^{0.42\pm 0.03},~L_{\mathrm{H}\alpha} \propto r^{2.72\pm 0.04},~L_{\mathrm{H}\alpha} \propto \sigma^{4.18\pm 0.21}$, and $L_{\mathrm{H}\alpha} \propto M_\mathrm{Jeans}^{1.24\pm 0.05}$). We discuss these results in the context of the existing simulations of clump formation and evolution, with an emphasis on the processes that drive-up the turbulent motions in the interstellar medium. Our results indicate that while the turbulence of discs may have important implications for the size and luminosity of regions which form within them, the same processes govern their formation from high redshift to the current epoch.\\

\end{abstract}
\begin{keywords}
galaxies: formation -- galaxies: high-redshift -- galaxies: star clusters -- galaxies: kinematics and dynamics -- ISM: H II regions.
\end{keywords}

\section{Introduction}

Star-forming galaxies at high redshift are generally clumpy, gas-rich, and highly efficient at forming stars \citep{2010ApJ...713..686D,2010Natur.463..781T,2011A&A...528A.124C}.
Compared to local star-forming galaxies that contain on average hundreds of \HII regions, galaxies in the early universe are made up of a handful of kilo-parsec sized star-forming `clumps' \citep{cowie:1995:10,van-den-bergh:1996:08}. These morphologies cannot be explained by simple resolution effects or an imaging K-correction \citep{2000bgfp.conf..257D}.  Artificially redshifting images of local star-forming galaxies with distributions of hundreds of \HII regions produces high-redshift images with mostly even light distributions \citep{2009ApJ...701..306E}. Furthermore, clumpy structures have been seen in broadband images as well as UV emission at high redshift, providing evidence for the build-up of massive star clusters \citep{2010arXiv1011.1507F,2011arXiv1110.3800G}. 

New kinematic data from integral field spectroscopy has supplemented the morphological studies to reveal that the clumps are embedded in massive rotating discs with high gas velocity dispersions of $50-200$ km s$^{-1}$ (e.g. \citealt{2008A&A...488...99V,Swinbank:2009tw,2009ApJ...706.1364F,Jones:2010uf,2011ApJ...733..101G, 2011MNRAS.417.2601W}). The clumps are found to have comparable velocity dispersions to the global disc dispersion with typical sizes of $\sim$1 kpc and masses of $\sim10^8-10^{10}$ \Msun~\citep{2005ApJ...627..632E, 2009ApJ...701..306E,2011ApJ...733..101G,2011ApJ...739...45F,2011arXiv1110.3800G}.

The kinematic observations are in broad agreement with models of unstable disc formation. In these models, motions in the turbulent interstellar medium (ISM) result in high local velocity dispersions that drive up the Jeans mass and Jeans length resulting in massive star-forming regions in gravitationally unstable discs. The timescale of the clumpy phase is predicted to be short, of order $\sim0.5$ Gyr.  The interactions of the clumps and high turbulence heats the gas resulting in a thick disc with scale height of $\sim1$ kpc. The more massive clumps then spiral to the centre of the galaxy due to dynamical friction to create a bulge \citep{1999ApJ...514...77N,2004A&A...413..547I,2004ApJ...611...20I,2007ApJ...670..237B,Elmegreen:2008fk}.  

In more recent models the turbulent discs are continuously fed by cold gas accretion which allows multiple episodes of instability. In these simulations, the clumpy phase can have a lifetime of a few Gyr. Individual clumps last for $\sim0.5$ Gyr if they have high enough star formation rates to survive destruction by stellar feedback \citep{Krumholz:2010fk}. Total star formation rates are approximately equal to the accretion rate with most of the galactic star formation occurring in the clumps ($\sim100$ \sfrunits; \citealt{2009ApJ...703..785D, 2010MNRAS.404.2151C}). The mass of individual kpc-size clumps, $\sim10^8-10^{10}$ \Msun, are a few per cent of the mass of the disc amounting to a total of 20\% of the mass of the disc \citep{2009ApJ...703..785D,Krumholz:2010fk,Genel:2010vn}. 

However, there is disagreement amongst the models as to how clumps are able to survive long enough to be prevalent in observations. Clumps with supersonic velocity dispersion and high star formation rates are expected to expend all their gas within a disc dynamical time (50 Myr; \citealt{2009ApJ...703..785D}) unless something supports them from collapse. Turbulence, a natural candidate, is needed to maintain the high local velocity dispersions and thus the large instability scale. However, the restoring force of turbulence is uncertain with possible sources including gravitational motions of clumps within the disc \citep{2009ApJ...703..785D,Krumholz:2010fk,2010ApJ...719.1230A}, disc self-gravity \citep{2004ApJ...611...20I,2007ApJ...670..237B,2010MNRAS.404.2151C,2010ApJ...725.2324B}, radiative feedback \citep{2010ApJ...709..191M,Krumholz:2010fk}, supernovae \citep{2004RvMP...76..125M,2006ApJ...653.1266J}, and gas accretion \citep{2006ApJ...645.1062F,2006Natur.442..786G,2010ApJ...719..229G}. However, turbulence is not the only mechanism for stabilising the clumps. If the regions become rotationally bound they may be supported against total free-fall collapse by their rotation \citep{2011arXiv1106.5587C,2009MNRAS.397L..64A,2010ApJ...719.1230A}. 

Furthermore, it is unknown how long individual clumps survive and how they reach their end. In the extreme environments expected in unstable gaseous discs, radiation pressure and momentum-driven galactic winds are expected and may be strong enough to disrupt the clumps before they are able to virialise and fall to the centre  to form a bulge via dynamical friction \citep{2010ApJ...709..191M,Genel:2010vn}.

Kinematic measurements of clump motions and turbulence are therefore essential for understanding how high-redshift galaxies form. Measurements of velocity dispersion of clumps have become more available with the advent of integral field spectroscopy (IFS) and adaptive optics (AO) systems \citep{Swinbank:2009tw, Jones:2010uf, 2011ApJ...733..101G}. These can be used to test clump formation models and the driving forces behind their survival by using gas velocity dispersion as a proxy for turbulence. 

Recent IFS observations have suggested that high-redshift clumps may form differently from local \HII regions as a result of a more turbulent environment with higher star-formation efficiencies, and gas densities. Clumps in intrinsically L$_{*}$ and sub-L$_{*}$ galaxies at $z>2$ have been found to be two orders of magnitude more luminous than \HII regions at a given radius \citep{Swinbank:2009tw,Jones:2010uf}, and their clump gas velocity dispersions lie in the range $60-100$ km s$^{-1}$, $2-6\times$ greater than local \HII regions \citep{2011ApJ...733..101G}. However, these properties are comparable to those of the largest \HII regions locally, commonly called `giant' \HII regions such as 30Doradus and II Zw 40 which have the same offset in star formation from `normal'-size \HII regions as high-redshift clumps. This offset in luminosity indicates that both classes have higher star formation surface densities than regular \HII regions suggesting that the most apt comparison at the different epochs is between clumps and giant \HII regions. 

In addition to \HII regions and giant \HII regions, \halpha results of high-redshift clumps have also been directly compared to local giant molecular clouds (GMCs; \citealt{2010ApJ...709..191M}). Some simulations are based on the assumption that giant clumps at high redshift represent single star-forming molecular clouds \citep{Krumholz:2010fk,2010ApJ...709..191M}. However, only a few clumps have been observed at high redshift from molecular gas by observing the regions in highly lensed galaxies \citep{2010Natur.464..733S}. It remains unclear what are the best analogues of high-redshift clumps at low redshift and whether the clumps form under the same conditions as GMCs or \HII regions. 

One strategy to compare star-forming regions at high and low redshifts is to look at scaling relations to see if they evolve with the cosmic time. An investigation of this kind is justified by the literature as extensive research has been dedicated to the relationships found between three key properties: luminosity, size, and velocity dispersion for giant \HII regions and GMCs (e.g. \citealt{1981MNRAS.195..839T, 1983ApJ...274..141G,1981MNRAS.194..809L,1988A&A...201..199A,2006A&A...455..539R,2006A&A...445..471B,2007A&A...472..421M}). The relationships between size, luminosity, and velocity dispersion have been greatly debated with proportionalities ranging from $\sigma \propto r^{1.14-3.68}$, $L\propto r^{1.92-3}$, $L \propto \sigma^{2.6-6.6}$ \citep{2000AJ....120..752F,2011AJ....141..113G}. However, these discussions often suffer from small-number statistics and biased measurements. We will show that, by bringing together multiple studies from the literature, measured in a self-consistent way, the spectrum of \HII regions form relatively consistent relationships between size, luminosity, and velocity dispersion, which can be extended to include high-redshift clumps. The scaling relationships observed have implications for how star-forming regions form in the changing cosmic environment.

\begin{table*}
\begin{minipage}{\textwidth}
\caption{High-Redshift Clump Samples}
\begin{tabular*}{\textwidth}{@{\extracolsep{\fill}}lclccl}
\hline
{Paper} & {$z$} & {Parent }&{Number of } & {Number of} & {Radius} \\
{} & {} & {Sample} &{Clumps} & {Galaxies} &{Method$^{a}$} \\
\hline
This Work  &  1.28$-$1.46 & WiggleZ & 8  &  3 & core \\
Jones et al. 2010  &   1.68$-$2.65 & lensed & 8  & 4 & isophote \\
Genzel et al. 2011 &  2.18$-$2.26 &   SINS & 5  & 3 & core \\
F\"{o}rster-Schreiber et al. 2011 &   2.26 &   SINS & 7  &  1 & photometric \\
Swinbank et al. 2009   & 4.92 & lensed & 2 & 1 &  isophote \\
\hline
\label{table.highz}
\end{tabular*}\\
$^{a}$ A more detailed description of how regions were measured is detailed in the Appendix.\\
\end{minipage}
\end{table*}

\begin{table*}
\begin{minipage}{\textwidth}
\caption{Low-Redshift \HII Region Samples}
\begin{tabular*}{\textwidth}{@{\extracolsep{\fill}}lllcll}
\hline
{Paper} & {Number of Galaxies} & {Parent Galaxy} & {N$^{a}$} & {Radius} & {Dispersion}\\
{} & {in Parent Sample} & {Types} & & {Method} & {Method}\\
\hline
SINGS (Kennicutt et al. 2003)  &   7 NGC & Spirals, Irregulars &  2091 &  core & $\ldots$  \\
Gallagher \& Hunter 1983   & 7 NGC; Mk 35;   & Irregulars & 30  &  core & echelle\\
&  Haro 22; A1004+10   &   & &  &\\
Arsenault et al. 1988  &  26 NGC; 5 IC; 5 HO & Spirals, Irregulars, & 57  &  isophotal & Fabry-Perot \\
&     &  Dwarf irregulars &  & \\
Bastian et al. 2006 &  NGC 4038/39 (Antennae) & LIRG & 4 & isophotal  & IFS \\
Rozas et al. 2006  &  10 NGC & Isolated spirals &43  &  40\% isophotal$^{b}$ & echelle \\
Monreal-Ibero et al. 2007 &  5 IRAS   & ULIRGs & 12 & half-light  & IFS \\
\hline
\label{table.lowz}
\end{tabular*}\\
$^{a}$ Number of \HII regions used in this paper.\\
$^{b}$ A method that measures core light. The size of the 40\% isophote is equal to the diameter of the circle containing the same area as 40\% of the peak region surface brightness. \\
\end{minipage}
\end{table*}

In this paper we present properties of eight clumps from three $z\sim1.3$ galaxies in the WiggleZ survey. These galaxies were selected from \cite{2011MNRAS.417.2601W}, herein Paper I, in which we presented \halpha kinematics of 13 star-forming galaxies observed with IFS. In Section 2 we introduce the WiggleZ clumps and a comparison sample at low and high redshift.  We address the most reproducible measures of size, luminosity, and velocity dispersion for comparison across 12 Gyrs ($z=0-5$) in Section 3 and apply them to the WiggleZ clumps. In Section 4 we quantify disc instabilities and estimate the mass of star-forming regions. In Section 5 we investigate the scaling relations of size, luminosity, and velocity dispersion of star-forming regions and their possible theoretical drivers. In Section 6 we use the results of Sections 3$-$5 to test predictions from models and simulations of clump formation. In Section 7 we summarise the properties of the WiggleZ clumps presented here. A standard $\Lambda$CDM cosmology of $\Omega_{\mathrm{m}} = 0.27$, $\Omega_{\Lambda} = 0.73$, $h = 0.7$ is adopted throughout this paper. In this cosmology, at redshift $z=1.3$, 1 arcsec corresponds to 8.6 kpc in physical coordinates.

\section{Data}
In this Section we present 8 new clumps, at $z>1$, in galaxies selected from the WiggleZ Dark Energy Survey \citep{Drinkwater:2010bx} and bring together comparison samples of star-forming regions at high and low redshift. For the comparison samples we only consider clump and \HII region data from \halpha observations for consistency and due to the wealth of data available in the literature. We focus on studies with published \halpha fluxes, sizes, and velocity dispersions for individual regions. A summary of the comparison data is given in Table \ref{table.highz} (high redshift) and Table~\ref{table.lowz} (low redshift), corrections to the distances, sizes, and cosmology that we introduce to improve the comparison of different samples are detailed in the Appendix.

\subsection{WiggleZ Clumps at $z\sim1.3$}
The new clump data presented in this paper are derived from the IFS observations of \halpha luminous galaxies at $z\sim1.3$ first introduced in Paper I. Data was taken with OSIRIS (OH Suppressing InfraRed Imaging Spectrograph; \citealt{2006NewAR..50..362L}) with the Keck II Laser guide star adaptive optics system \citep{2006PASP..118..297W, 2006PASP..118..310V}. OSIRIS is a lenslet array spectrograph with a $2048 \times 2048$ Hawaii-2 detector and spectral resolution $R\sim3600$.  All galaxies were observed with the 0.05$''$ pixel scale.  The FWHM of the tip-tilt stars ranged from one and a half to two pixels, 0.062-0.1$''$, with an average Strehl ratio of 30\%  estimated from the tip-tilt stars.

Of the 13 galaxies presented in Paper I, three galaxies have \halpha morphologies with multiple resolved star-forming regions. A summary of the global properties of these galaxies are given in Table~\ref{tab.global}. From this data we measure individual \halpha clumps and their kinematic properties. Clumps are identified solely from 2D \halpha emission images. The unsmoothed raw \halpha images used throughout this analysis were created from a summation over the spectral dimension within 1-sigma of the systemic velocity, as determined by a single Gaussian fit to the integrated spectrum. None of the galaxies have detectable continuum that could contaminate the \halpha images. We find 8 resolved clumps in total. The identification of clumps and measurement of their properties are discussed in Section 3. 

\begin{table*}
\begin{minipage}{\textwidth}
\caption{Global Properties of WiggleZ Clumpy Galaxies}
\begin{tabular*}{\textwidth}{@{\extracolsep{\fill}}lcccccccc}
\hline
{ID	} & 
{$z$ } & 
{$f_{H\alpha}^{a}$} & 
{log[M$_{*}$(\Msun)] } &  
{\smean } &
{$\sigma_\mathrm{net}$} &
{\vshear } & 
{$12+log$(O/H)} &
{N$^{b}$ } \\
{} & 
{} & 
{} & 
{} &
{(km s$^{-1}$)} & 
{(km s$^{-1}$)} & 
{(km s$^{-1}$)} & 
{} &
{}  \\
\hline
WK0912\_13R &  1.2873 & 48    $\pm$ 16.1  & 10.7 $\pm$ 0.2   & 81.6 $\pm$ 27.7 & 104.3 $\pm$ 27.1 & 130.5 $\pm$ 11.0 &    $<$8.47 	   & 2 \\
WK1002\_61S &  1.3039 & 81.5  $\pm$ 5.3   & 10.7 $\pm$ 0.7  & 85.1 $\pm$ 20.3  & 88.4 $\pm$ 4.4 & 71.5 $\pm$ 10.7   &    $<$8.37      & 3 \\
WK0909\_06R &  1.4602 & 94.9  $\pm$ 6.7   & 11.0 $\pm$ 0.2  & 92.6 $\pm$ 27.8  & 153.3 $\pm$ 8.4 & 160.8 $\pm$ 2.4   & 8.55 $\pm$ 0.1  & 3 \\
\hline
\label{tab.global}
\end{tabular*}\\
$^{a}$Emission line flux in units of 10$^{-17}$ ergs~s$^{-1}$~cm$^{-2}$. \\
$^{b}$Number of clumps measured in each galaxy.
\end{minipage}
\end{table*}

\subsection{High-redshift comparison sample}
At high redshift the study of clumps suffers from low-number statistics due to the resolution limits of current instrumentation from the ground. The high-redshift comparison sample consists of 22 clumps at $z>1.6$ as given in Table~\ref{table.highz}, which are all at higher redshift than the WiggleZ clumps. The data for two regions are taken from \cite{Swinbank:2009tw} for the lensed $z=4.92$ galaxy Cl1358+62. Eight regions are obtained from the \cite{Jones:2010uf} lensed sample at $z\sim2$, and twelve regions are taken from the SINS survey \citep{2011ApJ...733..101G,2011ApJ...739...45F} at $z\sim2$. We also compare to simulated high-redshift ($z\sim1.9-3.0$) clumps extracted from five simulated gravitationally unstable discs with baryonic masses of $10^{10}-2\times10^{11}$~\Msun~at different stages of their lifetime \citep{2011arXiv1106.5587C}. 

\vspace*{-0.1cm}
\subsection{Low-redshift comparison sample}
At low redshift we select a sample of extragalactic \HII regions and a sample of giant extragalactic \HII regions. The difference between `normal' \HII regions and `giant' \HII regions is determined not only by greater size and luminosity but primarily by having supersonic turbulence ($\sigma>c_\mathrm{sound}$), a property that was later integrated into the definition of giant \HII regions \citep{1970ApJ...161...33S,1981MNRAS.195..839T,1983ApJ...274..141G}. The sound velocity for \HII regions is defined as $c_\mathrm{\HII} = \sqrt{k\mu mT_\mathrm{\HII}} = 13$ km s$^{-1}$ where $k$ is the Boltzmann constant, $\mu$ is the mean molecular weight, m is the mass of hydrogen, and $T_\mathrm{\HII}$ is the temperature for \HII regions ($\sim10^4$ K; \citealt{2000AJ....120..752F}). For the velocity dispersions to be supersonic a mechanism is required to maintain the turbulence, which is expected to dissipate by shocks within a free-fall time, making them interesting objects to study. 

The comparison data for the normal \HII regions are derived from the Spitzer Infrared Nearby Galaxies Survey (SINGS, \citealt{2003PASP..115..928K}). The size and luminosity of this sample are measured from publicly-available $R$-band continuum subtracted \halpha images using the same methods applied to the WiggleZ data as described in Section 3.  We identified 2091 regions from a total of 7 galaxies. The Extragalactic Distance Database (EDD, \citealt{2009AJ....138..323T}) is used for consistent distance indicators for the SINGS galaxies. 

The data for giant \HII regions are taken from the literature from different parent populations including ULIRGS, dwarf irregulars, and isolated spirals \citep{1983ApJ...274..141G,1988A&A...201..199A,2006A&A...455..539R,2006A&A...445..471B,2007A&A...472..421M}. Studies that included \halpha velocity dispersion were preferentially chosen from the literature for comparison to the high-redshift IFS data. Values are corrected to the distances from the EDD and cosmology defined in Section 1. Corrections to published data that we introduce to improve the comparison of different samples are described in detail in the Appendix. 

\section{Region parameters}
To accurately compare the data from the WiggleZ clumps with the two comparison samples the clump properties from each sample should be measured in a consistent way. In this section we investigate the best methods to determine the radius, luminosity, velocity dispersion, and metallicity, and apply them to local \HII regions in the SINGS galaxies and to $z\sim1.3$ clumps in the WiggleZ galaxies. We consider the effects of resolution on these measurements and the implications they will have for the scaling relations studied in Section 5. Results are summarised in Table~\ref{tab.clumps}.  

\subsection{Region size}

\subsubsection{How to measure size: isophote vs. profile fitting?}
The size of an individual star-forming region is a difficult but important physical property to accurately measure. Subsequent calculations that depend on the size include region luminosity, velocity dispersion, relative velocity, and metallicity. Two methods are traditionally employed to measure region size: the isophote method \citep{1974ApJ...190..525S} and profile fitting, historically called the `core' method \citep{1979ApJ...228..696K}. In the isophote method a radius is derived from the area enclosing a total flux greater than a defined fraction of light or surface brightness level. The isophotal regions selected can be any shape with the radius then characterised from the total area assuming circular symmetry. Depending on the isophote level selected, the regions identified can represent either the cores of \HII regions or the core and surrounding diffuse nebulae. The standard choice of isophote varies among studies, with some isophotes selected to encompass a certain percentage of flux while others are selected visually. Conversely, the core method, or profile fitting, measures size by fitting a light profile, most commonly a Gaussian, to the surface brightness profile of each region. The core radius primarily measures the ionised central regions of \HII regions, above the surrounding diffuse nebulae. 

\begin{figure*}
\includegraphics[scale=0.70]{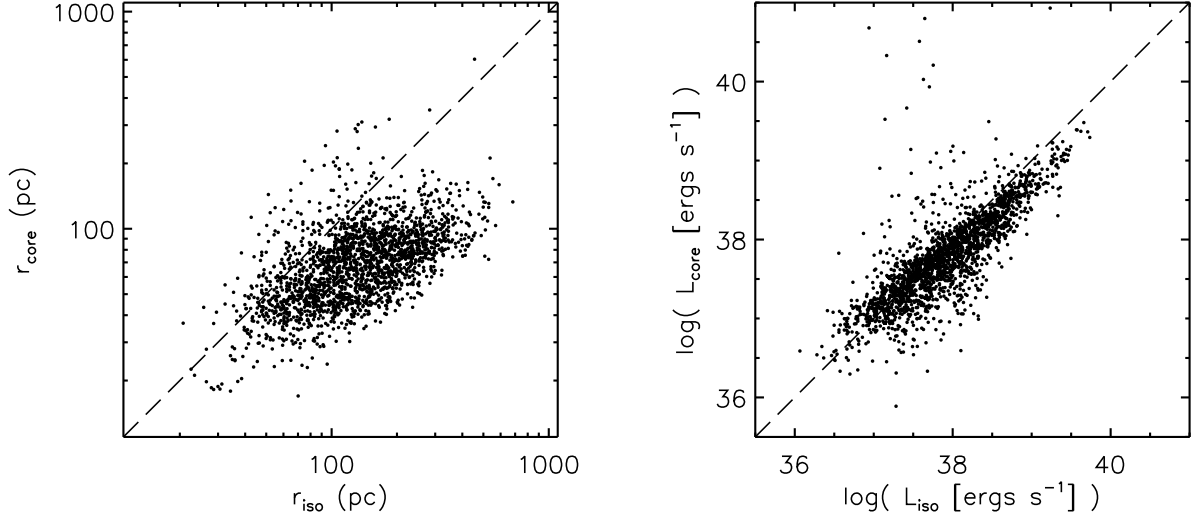}
\caption{Comparison between radii and luminosity of \HII regions measured from the isophote method and the core method in seven galaxies from the SINGS survey (NGC0024, NGC0628, NGC0925, NGC1566, NGC3938, NGC4254, and NGC7552).
We find that the core method measures systematically smaller \HII region sizes (left panel) within a small range of sizes but the luminosity measured in both methods (right panel) is consistent. This discrepancy is due to the inclusion of diffuse emission included in some isophotal radii. Although diffuse emission can significantly inflate region sizes, region luminosity is dominated by core emission. Dashed black lines show one-to-one relations.}
\label{fig.cmp}
\end{figure*}

We measure the isophotal and core radii of \HII regions in seven galaxies from the SINGS survey (NGC0024, NGC0628, NGC0925, NGC1566, NGC3938, NGC4254, and NGC7552). The isophote method selects regions in the SINGS galaxies with surface brightness greater than $1\times10^{-16}$ ergs s$^{-1}$ cm$^{-2}$ arcsec$^{-2}$. This isophote was selected for the ability to distinguish between close bright \HII regions while still selecting the majority of \HII regions in the core, arms, and halo of the listed galaxies. Isophotal radii, $r_{\mathrm{iso}}$, were calculated from isophote areas assuming circular symmetry. The core sizes, $r_{\mathrm{core}}$, were measured by fitting a Gaussian to the 1D radial surface brightness profile of each \HII region, which provides a good approximation to the data. This profile is chosen because it is the commonly used in the literature data, and the signal-to-noise (S/N) of the data does not justify a more sophisticated fit. 

Fig.~\ref{fig.cmp} shows a comparison between radii measured from the isophote method and the core method for the SINGS galaxies. Whilst  the isophote method measures a larger range of sizes (left panel), the luminosity measured in both methods (right panel) is approximately consistent.  The discrepancy in measurements arises from how each method accounts for the \halpha background, which can affect both size and flux measurements. Small errors, or deviations, in background levels produce large errors in isophotal radii, especially where the diffuse \halpha background is bright. Local diffuse tails or halos of \halpha emission around \HII regions can also greatly increase the radii causing a significant increase in the scatter of measured sizes. Thus, both image quality and \HII region sub-structures can contribute to the discrepancy between sizes from the different measurement methods. Despite the effect of the local background on size, the isophote method yields little to no offset to higher luminosities because the \HII region luminosity is dominated by the core. In the core method the total flux of the core is measured above the best fit local background level, a free parameter in the Gaussian fit, and therefore the core size measurement contains minimal bias due to the background emission. 

The issue of background becomes particularly salient at high redshift. Diffuse \HII halos are either undetected due to their low surface brightness or are smeared together increasing the local background. Therefore, at high redshift the local background could become a significant contaminant, increasing luminosity at a given size as a result of poorer resolution. This would also be an issue when regions are close together, separated by just a few pixels, causing clump light to overlap and combine.

We argue that the core radius is an observationally better determined parameter. The boundary of a core is better defined physically and is less likely to be contaminated by the local diffuse nebular gas. No surface brightness level needs to be assumed or tuned to select \HII regions. This method also better serves the comparison of low redshift and high redshift data as no flux level or signal-to-noise is required for direct comparison. This could be particularly important for high-redshift data in which clumps are only marginally resolved and often separated by only a few pixels. Finally, $r_\mathrm{core}$ is better justified theoretically because it more closely approximates the central ionised cluster, which will become relevant for future discussions of what drives the relations between radius and and other observed parameters (Section 5). A disadvantage of the core method is that it assumes the light profile is a 2D Gaussian and will be subject to systematic error for other distributions.
\begin{table*}
\begin{minipage}{\textwidth}
\caption{Properties of Clumps in WiggleZ galaxies from OSIRIS \halpha datacube}
\begin{tabular*}{\textwidth}{@{\extracolsep{\fill}}lccccccc}
\hline
{ID	} & 
{$v_{rel}^{a}$ } & 
{$\sigma_{net}^{b}$ } &  
{SFR$_{\mathrm{H}\alpha}^{b}$	   } & 
{$r_{core}$ } & 
{$r_{iso}$ } & 
{log[M$_{*}$(\Msun)] } &
{$12+log$(O/H)$^{b}$ } \\
{} & 
{(km s$^{-1}$)	} & 
{(km s$^{-1}$)	} & 
{(\sfrunits)	} & 
{(pc)} & 
{(pc)} & 
{} &
{ }  \\
\hline
WK0912\_13R & \ldots & 88.4 $\pm$ 4.4 &  20.3 $\pm$ 6.8 & \ldots & \ldots & \ldots &  $<$8.47 \\
clump-1 &  55.4 $\pm$ 67.4  & 140.5 $\pm$ 66.5 & 2.80 $\pm$ 1.87  & 1400 & 1800 & 10.5  & $<$8.76          \\
clump-2 & -39.5 $\pm$ 18.1  & 52.03 $\pm$ 18.1 & 1.44 $\pm$ 0.64  & 1000 & 1100 & 9.5   & $<$8.60          \\
 \\
WK1002\_61S & \ldots & 104.3 $\pm$ 27.7& 35.6 $\pm$ 2.3 & \ldots & \ldots & \ldots & $<$8.37 \\
clump-1 & 27.5 $\pm$ 6.4   & 77.65 $\pm$ 6.41 & 8.31 $\pm$ 0.89  & 1600 & 1800 & 10.0  & $<$8.42          \\
clump-2 & 3.2  $\pm$ 11.9  & 90.74 $\pm$ 11.9 & 2.54 $\pm$ 0.43  & 900  & 1200 & 10.0  & $<$8.46          \\
clump-3 &-32.7 $\pm$ 22.9  & 113.5 $\pm$ 22.8 & 4.98 $\pm$ 1.32  & 1700 & 700  & 10.4  & $<$8.62          \\
\\
WK0909\_06R & \ldots & 153.3 $\pm$ 8.4  & 54.8 $\pm$ 3.9 & \ldots & \ldots & \ldots & 8.55 $\pm$0.10 \\
clump-1 & 130.7 $\pm$ 7.5   & 135.6 $\pm$ 7.48 & 6.62 $\pm$ 0.49  & 1200 & 2100 & 10.4  & 8.57 $\pm$ 0.10  \\
clump-2 & -45.7 $\pm$ 6.2   & 72.68 $\pm$ 6.23 & 2.13 $\pm$ 0.24  & 800  & 1500 & 9.7   & 8.38 $\pm$ 0.29  \\
clump-3 & -88.3 $\pm$ 6.4   & 80.02 $\pm$ 6.43 & 11.3 $\pm$ 1.2   & 3000 & 1100 & 10.3  & 8.44 $\pm$ 0.21  \\
\hline
\label{tab.clumps}
\end{tabular*}\\
$^{a}$Velocity relative to the systematic redshift.\\
$^{b}$Calculated within $1~r_{core}$.\\
\vspace{-1cm}
\end{minipage}
\end{table*}

\begin{figure*}
\includegraphics[scale=0.9]{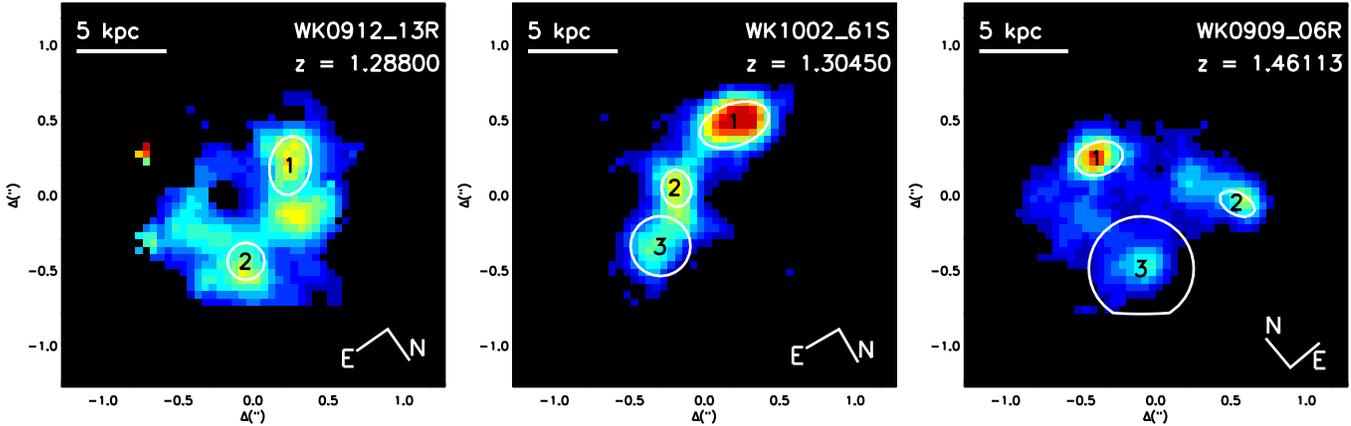}
\vspace{-1.5 cm}
\caption{Smoothed \halpha flux maps of three clumpy galaxies in the WiggleZ kinematic sample observed with the OSIRIS integral field spectrograph. Clumps identified in Section 3 are shown by the white ellipses. Pixel size corresponds to 0.05$''$ and the physical orientation are denoted by the compasses in the bottom right corner. Images are smoothed with a Gaussian kernel with FWHM = 0.15$''$. }
\label{fig.clumps}
\end{figure*}

\subsubsection{WiggleZ sizes}
Turning now to the WiggleZ sample, we measure the core and isophotal radii of 8 clumps from the 3 galaxies which show multiple resolved clumps, WK0912\_13R, WK1002\_61S, WK0909\_06R. Core radii of the WiggleZ clumps were measured by fitting multiple 2D Gaussians simultaneously to the unsmoothed \halpha images. The measurements given in Table~\ref{tab.clumps} represent the 1-sigma Gaussian widths. In cases were clumps are elliptical, an average radius was calculated. To measure the isophotal radii the isophotes used, $4.5-6.5\times10^{-15}$ ergs s$^{-1}$ cm$^{-2}$ arcsec$^{-2}$, were tuned to obtain a reasonable comparison of $r_\mathrm{core}$ and $r_\mathrm{iso}$. Both methods identified the same regions with comparable luminosities. Measurements of $r_\mathrm{core}$ and $r_\mathrm{iso}$ are given in Table~\ref{tab.clumps} and the identified clumps are shown in Fig.~\ref{fig.clumps} with the core region sizes overplotted on the \halpha images. In the remainder of this paper we use the core radii of the WiggleZ clumps to represent their sizes, $r_\mathrm{cl}\equiv r_\mathrm{core}$.

It is likely that the measured clump radius over-estimates the actual size of clump-3 in object WK0909\_06R. Admittedly this represents a failing of the core method, such that low S/N diffuse regions are difficult to model with a Gaussian. In object WK0912\_13R three clumps can be identified visually in the smoothed image (Fig.~\ref{fig.clumps}). However, when fitting 2D Gaussians to the raw images the addition of a third is not justified by the data as this region is unresolved. This region is also not identified at the selected isophote level, which yields similar sizes for clump-1 and clump-2 from the core and isophote methods. Given the uncertainty in measuring clump sizes, arising from degeneracies in the 2D Gaussian fits, resolution effects, and systematic error introduced if the regions are not truly Gaussian, we estimate a $\sim 30$\% error on the sizes.

\subsubsection{Resolution effects}
The 8 WiggleZ clumps presented here are resolved as are the star-forming regions in the comparison samples. However, it is possible that some WiggleZ clumps are aggregates of unresolved regions since not all clumps are spherical. The resolution of the OSIRIS \halpha images studied in the WiggleZ sample is $\sim520-840$ pc in physical coordinates at $z=1.3$. Because the WiggleZ clumps are close to being point sources, the \halpha luminosity may be underestimated due to the Strehl ratio ($\sim30$\% estimated from the tip-tilt stars) as a result of light being spread out by the uncorrected seeing-limited halo.

Despite the regions being resolved, the image resolution may still have implications for the scaling relations investigate in Section 5. Resolution can have the following effects: (1) a core and diffuse emission are smeared together, (2) multiple core regions are smeared together, (3) diffuse emission is smeared to resemble a core, and (4) small regions are reduced to noise, with all effects resulting in inflated size and luminosity measurements for the regions observed \citep{2000A&A...361..913P}. \cite{Swinbank:2009tw} investigate the effects of redshifting local galaxy images to $z>2$ and degrading the resolution on the size and luminosity of regions. They find that only the cores of the highest luminosity \HII regions could be observed at high redshift and that lowering the resolution of images blends multiple regions into larger more luminous regions. Both studies find that degrading the resolution results in a steepening of the empirically measured slope between size and luminosity. We readdress these resolution effects in Section 5.

\subsection{Region luminosity}
The luminosity of each WiggleZ clump is measured from the integrated spectrum created from the spaxels within the 1-sigma contour of the best Gaussian 2D fit.  A Gaussian profile was fitted to the \halpha emission in the clump integrated spectrum with the wavelength, intensity and width as free parameters. Errors were calculated by adding in quadrature an estimation of the sky noise from the variance of the spectrum offset from the emission lines with the Poisson error of the photon counting statistics in the emission lines. Clump star formation rates (SFR) range from $1-12$ \Msun~yr$^{-1}$ derived from the \halpha emission using the \cite{1998ARA&A..36..189K} conversion with a Baldry-Glazebrook initial mass function (IMF; \citealt{2003ApJ...593..258B}, BG03) and are given in Table~\ref{tab.clumps} with star formation rate densities, $\Sigma_*=\mathrm{SFR}/\pi r_\mathrm{core}^2\sim0.4-2$ \Msun~yr$^{-1}$ kpc$^{-2}$. All the local and high-redshift data have been converted to a BG03 IMF for comparisons in Section 5.  A division by 1.82 is required to convert a SFR with a Salpter IMF to a SFR with a BG03 IMF.

For {WK0912\_13R}, {WK1002\_61S}, and {WK0909\_06R} the clumps make up 21\%, 45\%, and 37\% of the total star formation respectively with each clump contributing $\sim10$\%. This is in relative agreement with other IFS studies of emission lines \citep{2011ApJ...733..101G} and broadband studies \citep{2011ApJ...739...45F,2011arXiv1110.3800G}. In galaxies in the Hubble Deep Field at $1.5<z<2.5$ up to 50\% of star formation occurs in clumps with a $\sim$10\% contribution per clump \citep{2011arXiv1110.3800G}. In the hydrodynamic cosmological simulations of \cite{2009MNRAS.397L..64A} and \cite{2010MNRAS.404.2151C} the giant clumps account for about half the total star formation  at a given time and 20\% of the disc mass.  For comparison, \HII regions in \halpha can contribute 30-50\% of the gas in local spiral galaxies \citep{1996AJ....111.2265F, 2000A&A...361..913P} and up to 75\% with beam-smearing \citep{2000A&A...361..913P}. Thus the contribution from clump light measured in the WiggleZ sample agrees with high-redshift emission-line samples, well-resolved low-redshift samples, and simulated samples.

\subsection{Region velocity dispersion}
\label{sec.winds}
The velocity dispersion of the gas will be dominated by the core region of emission. Although radii measurements are critical for size-luminosity comparisons, variations in the size of the regions have little effect on the integrated velocity dispersion (e.g. \citealt{1988A&A...201..199A}). We test this by measuring the velocity dispersion from the integrated spectrum in the clumps of the WiggleZ galaxies within regions of increasing radii. We find that the velocity dispersion does not change by more than one standard deviation when the radius is increased by a factor of two. As such, although some velocity dispersions in the local sample \citep{1988A&A...201..199A} and the high-redshift sample \citep{Jones:2010uf,Swinbank:2009tw} are measured within isophotal radii, the effect on their velocity dispersion measurements should be minimal. 

Velocity dispersions, $\sigma_\mathrm{net}$, for the WiggleZ clumps are derived from the integrated clump spectrum and corrected for instrumental broadening by subtracting the OSIRIS resolution in quadrature and are given in Table~\ref{tab.clumps}. We use this measure of velocity dispersion rather than the flux-weighted velocity dispersion, $\sigma_\mathrm{mean}$ \citep{2007ApJ...669..929L,Green:2010fk} or other measurements \citep{2011ApJ...741...69D}, for better comparison to the values taken from the literature as it more closely resembles a long-slit observation. Rotation signatures across the clumps are measured to be $\sim20-80$ km s$^{-1}$, with average $\Delta v/\Delta r_{cl}\sim15$ km s$^{-1}$ kpc$^{-1}$, where $\Delta v$ is the change in velocity across the clump. This is a small effect in comparison with measured velocity dispersions, as such no correction has been applied. Velocity dispersions of the clumps, uncorrected for beam-smearing, are either consistent with or slightly lower than the total disc dispersions. Total disc dispersions, measured in Paper I, are given in Table~\ref{tab.global}. Recent simulations show that even with moderate beam smearing $-$ 0.1 arcsec $-$ the actual observed clump dispersion is highly contaminated by the disc dispersion making it difficult to distinguish between the two signatures \citep{2011arXiv1106.5587C}. However, the same simulations predict that clump and disc dispersions should be physically comparable in the absence of beam-smearing, minimising the ramifications for our measurements.  

\begin{figure}
\includegraphics[scale=0.52, trim = 11mm 0mm 6mm 2mm, clip]{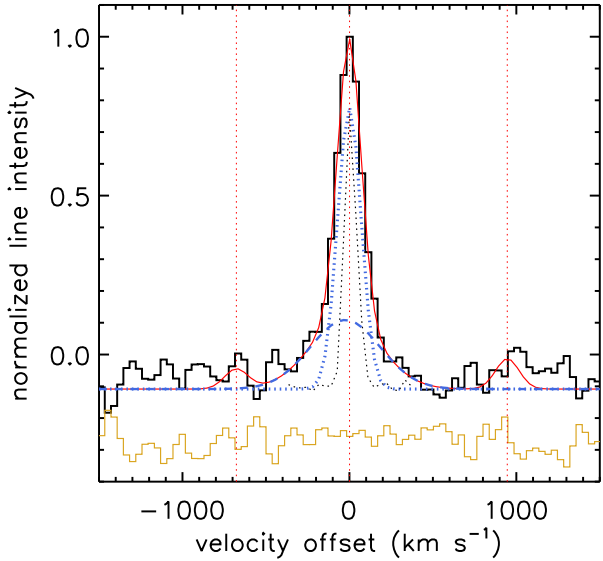}
\caption{Stack of the integrated \halpha spectra from 8 WiggleZ clumps corrected for velocity broadening. The red line is the best-fit model to the \halpha and \NII~emission lines composed of two \halpha Gaussian components. The blue dotted line is the narrow component with FWHM~$=170$ km s$^{-1}$ and the blue dashed line is the broad component with FWHM~$=490$ km s$^{-1}$ which has a consistent systemic velocity with relative offset $-24\pm30$ km s$^{-1}$. The broad component contains 40\% of the total flux. The orange line is the residual of the best-fit model subtracted from the data. The narrow black dotted line at zero velocity shows the instrumental resolution from a stack of sky lines in the observed spectra. Vertical dotted lines show the expected positions of \NII$\lambda6548$, \halphans$\lambda6563$, and \NII$\lambda6584$.}
\label{fig.winds}
\end{figure}

Clumps that have higher mass surface densities than the disc, $\Sigma_\mathrm{cl}>>\Sigma_\mathrm{d}$, could lead to higher dispersions within the clumps due to intense star formation or stellar winds. Large-scale winds can be launched from regions forming stars at $\Sigma_* >0.1$ \Msun~yr$^{-1}$ kpc$^{-2}$ \citep{2000ApJS..129..493H} and indeed the WiggleZ clumps measured here have star formation surface densities $\gtrsim10\times$ this threshold. If large-scale stellar outflows are driven by the star formation in clumps, inflated velocity dispersions are expected due to an underlying broad wind component \citep{1990ApJS...74..833H,2011ApJ...733..101G}. In Fig.~\ref{fig.winds} we stack the integrated spectrum from all 8 WiggleZ clumps at rest wavelength, removing any velocity broadening \citep{Shapiro:2009sj}. The stacked spectrum is fit by two Gaussian components, a narrow peak (FWHM $\sim 170$ km s$^{-1}$), and underlying broad peak (FWHM $\sim 490$ km s$^{-1}$) which has a consistent systemic velocity with relative offset $-24\pm30$ km s$^{-1}$. We find that a 2-Gaussian fit provides a significant improvement in the value of the chi-squared statistic compared to a 1-Gaussian fit, producing a decrease $\Delta \chi^2 \sim 60$ for the addition of 3 extra parameters and a reduced $\chi^2$ of 1.07. The broad Gaussian contributes 40\% to the total flux. We rule out AGN as the driver of the broad \halpha component because the clumps are off-centre from the photometric and kinematic centre of the galaxies \citep{Alexander:2010bd} and no evidence is seen on galactic scale for AGN contamination, as discussed in Paper I.

The broad velocity component observed in the stacked WiggleZ clumps has a smaller velocity width and smaller blueshift compared to $z\sim2$ starbursts galaxies, which may have broad components with up to FWHM $\sim1500$ km s$^{-1}$ blueshifted by 50 km s$^{-1}$ \citep{2011ApJ...733..101G,Tiran:2011fk}. However, we find that the broad velocity component of the WiggleZ clumps contributes a comparable or larger fraction to the overall flux, equal to 40\%. Although a broad component is not detected in the spectra of individual clumps, the presence of a significant broad component in the stack of clump spectra indicates that individual clump velocity dispersion measurements may be over-estimated due to contamination from large-scale winds. By comparing to a single Gaussian fit to the \halpha emission of the clump stack, this is estimated to be approximately a 15\% effect, or on average $\sim14$ km s$^{-1}$, comparable to the errors given in Table~\ref{tab.clumps}. The possible effect of winds on both clump luminosity and velocity dispersion is reflected in the error bars in the subsequent figures.

\subsection{Clump metallicity and ages}
It has been shown that the scatter in \HII region measurements can be partially explained by region metallicity \citep{1981MNRAS.195..839T,1987MNRAS.226..849M}. Variations in metallicity may also reveal details about the age of clumps, important to their evolution. We calculate clump metallicities for each individual WiggleZ clump and for the clump stack shown in Fig.~\ref{fig.winds} from the \NII/\halpha ratio following \cite{2004MNRAS.348L..59P};
\begin{eqnarray}
12+\mathrm{log} \left ( \frac{\mathrm{O}}{\mathrm{H}} \right ) = 8.90 +0.57 \times \mathrm{log} \left ( \frac{ \mathrm{\NII} }{ \mathrm{H}\alpha } \right ).
\label{eq.metallicity}
\end{eqnarray}
These values are given in Table~\ref{tab.clumps}. For individual WiggleZ clumps a single Gaussian is fit to the \halpha and \NII~emission lines. For the clump stack, the \halpha flux used in equation~(\ref{eq.metallicity}) is from the narrow velocity component (shown in Fig.~\ref{fig.winds} as the dotted blue line). The metallicity of all but the WK0909\_06R clumps are upper limits as \NII~in this sample is typically undetected. The stack of eight clumps has a metallicity of $12+\log(\mathrm{O/H})=8.35\pm0.31$. All metallicities in individual galaxies are consistent within the errors, which agrees with the simulations of \cite{2004A&A...413..547I} and \cite{2008A&A...486..741B} and implies a common clump evolution history.

\section{Disc instability}
\begin{figure*}
\includegraphics[scale=0.9]{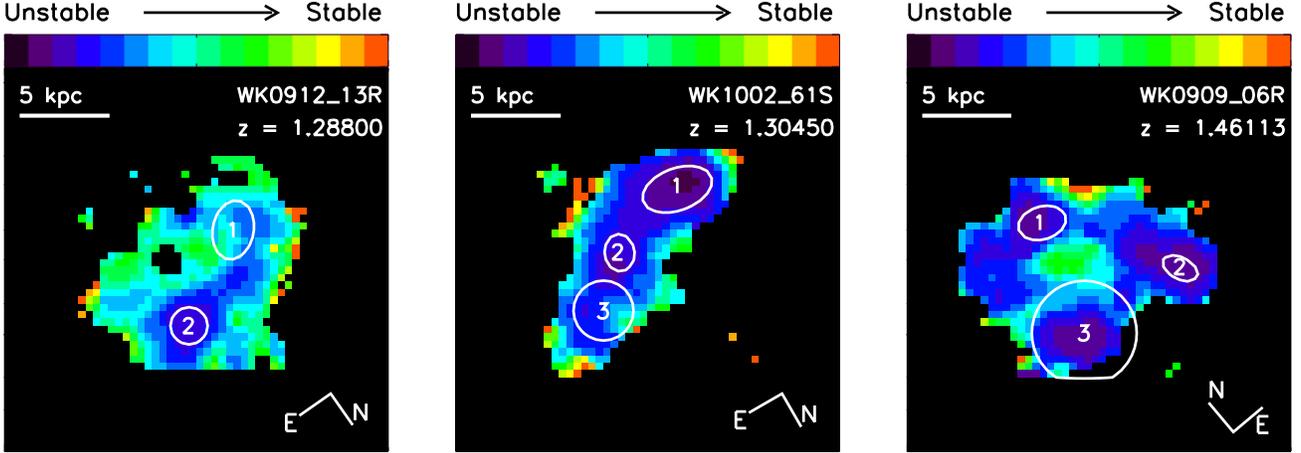}
\vspace{-1.5 cm}
\caption{Toomre parameter, $Q_\mathrm{gas}$ (equation~\ref{eq.toomre2}), mapped over three clumpy galaxies in the WiggleZ kinematic sample. The calculation of the Toomre parameter assumes the Kennicutt-Schmidt law and is uncertain due to the degeneracies of the variables in equation~(\ref{eq.toomre2}). As a result a relative linear scale for $Q_\mathrm{gas}$ is shown. All three WiggleZ galaxies have areas that are consistent with being in an unstable state ($Q_\mathrm{gas}$$<1$) within the errors. Clumps identified in Section 3 are shown by the white ellipses. Likely unstable regions are shown in purple and dark blue. The clumps correlate approximately with Toomre unstable regions.}
\label{fig.toomre}
\end{figure*}

As outlined in Section 1, models predict that clumps form out of gravitational instabilities in gas-rich turbulent discs. The high levels of turbulence are predicted to drive up the expected instability scale in disc galaxies resulting in massive star-forming regions. In this section we characterise the size and mass instability scale of the WiggleZ discs given the high velocity dispersions measured in Paper I and the previous section. We compare the WiggleZ clumps to the high-redshift and low-redshift samples, which have velocity dispersions ranging from $10-100$ km s$^{-1}$. We calculate relative Toomre parameters, shown in Fig.~\ref{fig.toomre}, to identify regions of instability where clumps could form. 

For star-forming regions to form in gaseous discs, the galaxies must become gravitationally unstable such that $Q<1$ where $Q$ is the Toomre parameter \citep{1964ApJ...139.1217T} defined for a pure gas disc as
\begin{eqnarray}
Q_{\mathrm{gas}} = \frac{\sigma_\mathrm{d}\kappa}{\pi\mathrm{G}\Sigma_{\mathrm{g}}}.
\label{eq.toomre1}
\end{eqnarray} 
To test these models of instability we calculate $Q_\mathrm{gas}$ for the three galaxies hosting clumps in the WiggleZ sample using
\begin{eqnarray}
Q_{\mathrm{gas}} = \frac{\sigma}{v_\mathrm{c}}\frac{a}{f_\mathrm{g}}
\label{eq.toomre2}
\end{eqnarray} 
as derived from \cite{2011ApJ...733..101G}, where $\kappa$ is the epicyclic frequency, $\sigma$ is the local disc gas velocity dispersion, $v_\mathrm{c}$ is the circular velocity, and $\Sigma_\mathrm{g}$ is the local gas surface density in the disc. The constant $a$ relates the epicyclic and circular frequency in different potentials such that  $a=\sqrt{2}$ for a disc with a flat rotation curve \citep{2008gady.book.....B}. We estimate the gas fraction, $f_\mathrm{g}$, from the inverse Kennicutt-Schmidt law (KS; \citealt{1959ApJ...129..243S,2007ApJ...671..333K}). 

In Paper I flat disc models were fit to the \halpha velocity maps of the WiggleZ galaxies. We obtained estimates for $v_\mathrm{c}\sin(i)$ rather than $v_\mathrm{c}$ because the inclination could not be constrained for lack of deep broadband data and the full extent of the kinematic gas disc is not observed. As a result the actual value of the Toomre parameter across the disc is highly uncertain. A relative Toomre parameter however is sufficient to identify the relative stability of the galaxy discs. We note that within the errors of $v_\mathrm{c}\sin(i)$ for the disc models, $Q_\mathrm{gas}$ for each galaxy has regions of instability ($Q_\mathrm{gas}<1$). In Fig.~\ref{fig.toomre} we show relative Toomre maps for {WK0912\_13R}, {WK1001\_61S}, and {WK0909\_06R}. Overlaid on the maps are the clumps identified in Section 3 from the core method. The clumps identified from the \halpha maps match to the less stable regions of the discs as expected from the models. These results are in agreement with IFS data at $z\sim2$ in which Toomre instability correlates with clump location \citep{2011ApJ...733..101G}. It is expected that low values of $Q_\mathrm{gas}$ will correlate with regions with high SFRs as the Toomre parameter is proportional to the inverse of star formation as a result of how the gas fraction is calculated. However, significant peaks in local velocity dispersion coincident with the clumps (for example if the clumps are virialised and rotationally supported) could counterbalance the local reduction of $Q_\mathrm{gas}$ by the star formation. However, this effect is not observed since significantly higher velocity dispersions are not measured in the clumps versus the disc.

\begin{figure}            
\vspace*{-0.3cm}\includegraphics[scale=0.52, trim = 11mm 0mm 6mm 2mm, clip]{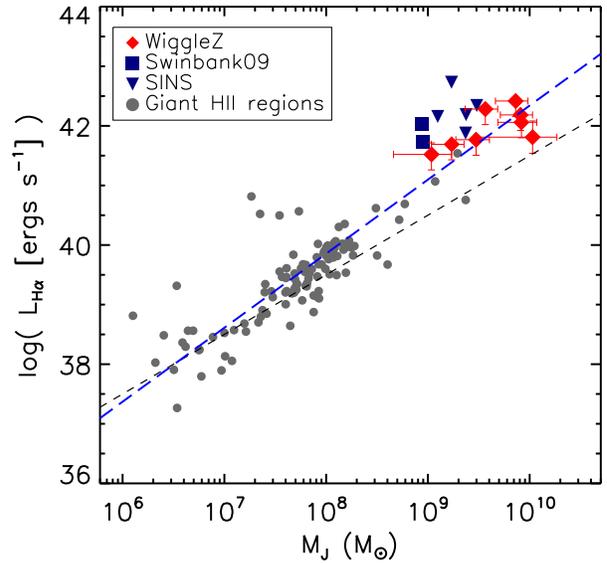}
\vspace*{-0.3cm}\caption{\halpha Luminosity vs. Jeans mass for the WiggleZ clumps (red diamonds), high-redshift comparison sample (blue filled symbols) and low-redshift giant \HII region sample (gray points). \halpha luminosity and Jeans mass ($M_\mathrm{J}=\pi^2 \sigma^2r_\mathrm{cl}/6\mathrm{G}$) show a tight correlation for star-forming regions in local star-forming galaxies (HII regions) to turbulent high-redshift unstable discs (clumps) with sizes of $\sim50-5000$ pc. The blue dashed line is the least squares best fit to the data, given by $L_{\mathrm{H}\alpha}[\mathrm{ergs~s}^{-1}] = 8\times10^{29}~M_\mathrm{J}[\mathrm{\Msun}]^{1.24\pm0.05}$. The black dot-dash line represents $L_{\mathrm{H}\alpha} \propto M_\mathrm{J}$ arbitrarily normalised for comparison. }
\vspace*{-0.7cm}\label{fig.mdyn}
\end{figure}

When discs are gravitationally unstable, clumps are expected to form under Jeans collapse on a characteristic scale. For the WiggleZ galaxies the expected masses and sizes of the clumps would be limited by the Jeans mass and length, given by
\begin{eqnarray}
\lambda_{\mathrm{J}} = \left ( \frac{\pi\sigma_\mathrm{d}^2}{\rho\mathrm{G}}\right )^{1/2},
\label{eq.jeans1}
\end{eqnarray}
and 
\begin{eqnarray}
M_{\mathrm{J}} = \frac{4\pi}{3} \left ( \frac{\lambda_\mathrm{J}}{2} \right )^{3} \rho,
\label{eq.jeans2}
\end{eqnarray}
where $\sigma_\mathrm{d}$ is the average velocity dispersion in the discs estimated by the velocity dispersions of the clumps and \HII regions,  and $\rho$ is the density of the disc. By combining equations (\ref{eq.jeans1}) and (\ref{eq.jeans2}) and assuming that the measured clump diameter is equal to a Jeans radius, we calculate Jeans masses for the WiggleZ clumps of $M_{\mathrm{J}}$ = $1.1\times10^9-1.1\times10^{10}$ \Msun, which are in agreement with the models (e.g. \citealt{Krumholz:2010fk}). In Fig.~\ref{fig.mdyn} we compare the Jeans mass to  \halpha luminosity for the WiggleZ clumps (red diamonds), the higher-redshift clumps (blue squares and triangles), and the low-redshift giant \HII regions (gray points). A tight continuous 
correlation is seen for all star-forming regions from $M_\mathrm{J}=10^6-10^{10}$ \Msun. The correlation in Fig.~\ref{fig.mdyn} combines the three main observables discussed in this paper with scatter characterised by a correlation coefficient of 0.90 and least squares best fit of $L_{\mathrm{H}\alpha} \propto M_\mathrm{J}^{1.24\pm0.05}$, indicating that the mass closely traces the ionised gas of star-forming regions.
\vspace*{-0.8cm}\begin{figure*}
\includegraphics[scale=0.85]{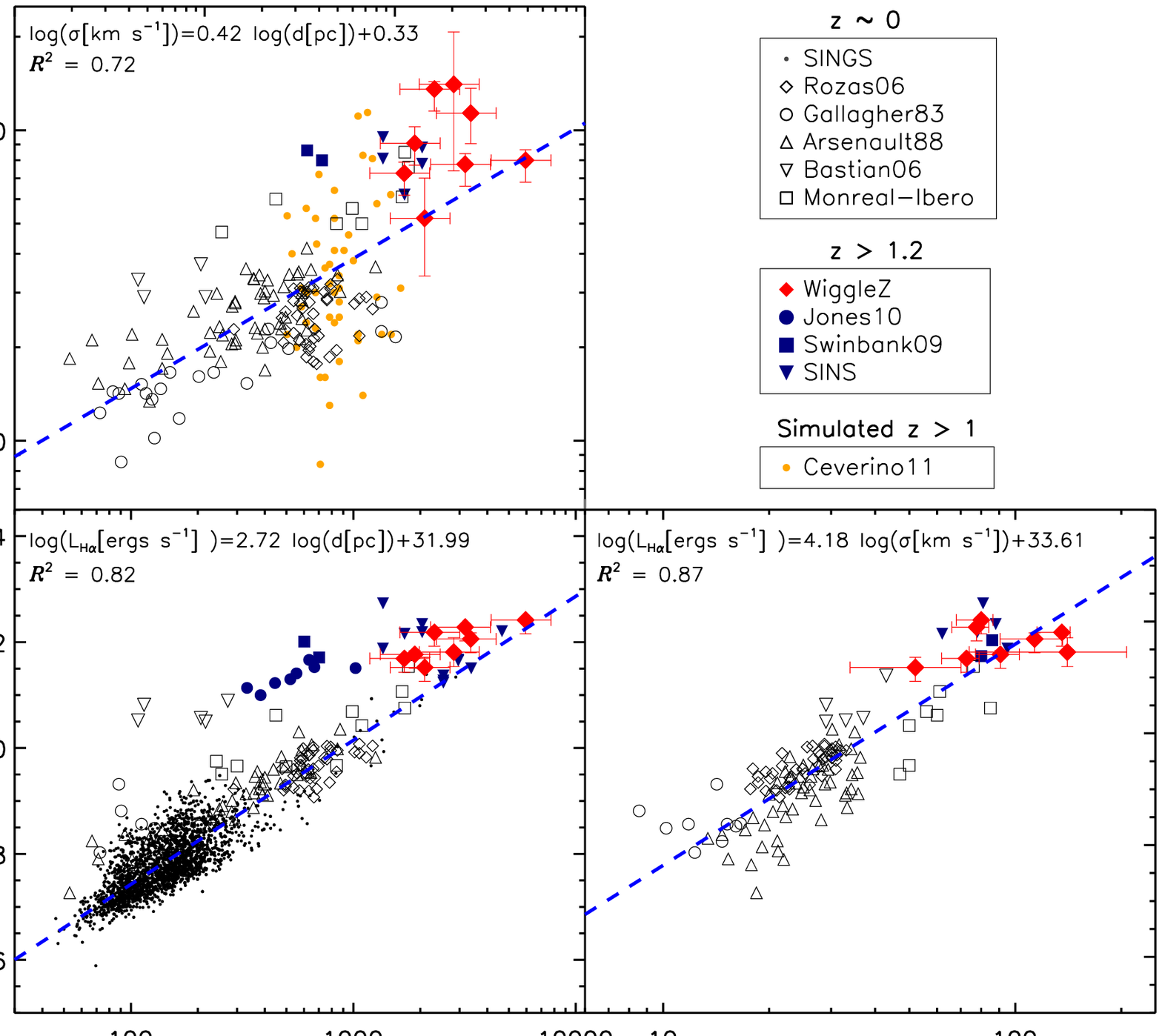}
\vspace{1.0 cm}
\caption{Relations between the \halpha luminosity, size, and velocity dispersion of \HII regions in local star-forming galaxies and clumps in high-redshift galaxies. Clumps from WiggleZ galaxies are represented by the bright red diamonds. High-redshift clumps are shown as filled symbols in blue and taken from \citet{Swinbank:2009tw,Jones:2010uf,2011ApJ...733..101G,2011ApJ...739...45F}.  \HII regions from local SINGS galaxies are displayed as black dots, giant \HII regions from local galaxies are shown as open black points and are taken from \citet{1983ApJ...274..141G,1988A&A...201..199A,2006A&A...445..471B,2006A&A...455..539R,2007A&A...472..421M}. The blue dashed lines show the least squares best fits, given in the top left corner of each panel with correlation coefficients, $R^2$, given directly below. Simulated clumps at $1.9<z<3.0$ from \citet{2011arXiv1106.5587C} are plotted in orange but have not been included in the fit.}
\label{fig.sizelumsigma}
\end{figure*}

\vspace*{0.2cm}\section{Scaling relations of star-forming regions}

Are the kpc-sized clumps found at high redshift formed out of disc instabilities in a different manner to local \HII regions, or are they more massive analogues as suggested by Fig.~\ref{fig.mdyn}? We show the correlations for measured \halpha luminosity, velocity dispersion, and size for the WiggleZ clumps, the high-redshift comparison sample, and the low-redshift comparison sample in Fig.~\ref{fig.sizelumsigma}. We find tight correlations between these properties over 50-5000 pc for \HII regions and clumps taken from eleven different studies. Dispersions range between $\sim10-100$ km s$^{-1}$ and \halpha luminosities span five orders of magnitude. Assuming equal weighting for all points we use linear least squares fitting to find the best fit to the relations to be 
\begin{eqnarray}
\hspace*{-0.7cm}&&\log(\sigma) =  (0.42\pm 0.03)\times\log(d) + ( 0.33\pm 0.09) , \nonumber\\     
\hspace*{-0.7cm}&&\log(L_{\mathrm{H}\alpha})=  (2.72\pm 0.04)\times\log(d) + (31.99\pm 0.08), \nonumber\\      
\hspace*{-0.7cm}&&\log(L_{\mathrm{H}\alpha})=  (4.18\pm 0.21)\times\log(\sigma) + (33.61\pm 0.31) , 
\end{eqnarray}
where $L_{\mathrm{H}\alpha}$ is measured in ergs s$^{-1}$, $d$ in pc, and the $\sigma$ in km s$^{-1}$, with correlation coefficients, $R^2$, of 0.72, 0.82, and 0.87 respectively. These fits are overplotted as blue dashed lines in Fig.~\ref{fig.sizelumsigma}. Simulated clumps are included in Fig.~\ref{fig.sizelumsigma} (orange points) for comparison to the data, but are not included in the above fits. Below we discuss the scaling relations and their possible drivers.

\subsection{Velocity Dispersion vs. Size}
We derive the relation between size and velocity dispersion for regions that form out of Jeans collapse in an isothermal disc. The disc scale height, $H$, for an isothermal disc is given by

\begin{eqnarray}
H  \equiv \frac{1}{\rho_{0}}\int_0^\infty \rho(z)~dz,
\end{eqnarray} 

\citep{1942ApJ....95..329S} where $\rho_0$  is the midplane density value and $z$ is the height above the plane. The scale length has the solution for a purely self-gravitating unmagnetised gas dominated disc

\begin{eqnarray}
H_\infty  = \frac{\sigma}{(2\pi \mathrm{G}\rho_\infty)^{1/2}} = \frac{\sigma^2}{\pi \mathrm{G} \Sigma_\mathrm{g}}, 
\end{eqnarray} 

where $H_\infty$ and $\rho_\infty$ are the disc scale height and density in this limit with $\sigma$ defined as the turbulent velocity dispersion. The surface density of the disc is given by $\Sigma_\mathrm{g}= 2H_\infty\rho_\infty$. Therefore we obtain an equation for the density such that

\begin{eqnarray}
\rho_\infty = \frac{\pi \mathrm{G} \Sigma_\mathrm{g}^2}{2\sigma^2}.
\end{eqnarray} 

Combining this with the definition of the Jeans length in equation~(\ref{eq.jeans1}), assuming $\rho = \rho_\infty$, we obtain a threshold length for the regions forming under Jeans collapse in an isothermal disc such that

\begin{eqnarray}
\lambda_\mathrm{J} < \frac{\sqrt{2}\sigma_\mathrm{d}^2 }{\mathrm{G}\Sigma_\mathrm{g}}.~
\label{eq.sigr}
\end{eqnarray} 

The corresponding mass limit from equation~(\ref{eq.jeans2}) becomes $M_\mathrm{J} < \sqrt{2}\pi^2 \sigma^4 / 6\mathrm{G}^2\Sigma_\mathrm{g}$.
To associate this scaling with measured size we adopt the convention of \cite{2009ApJ...703..785D} which relates the typical clump radius to the Jeans scale as $r_\mathrm{cl} \simeq \lambda_\mathrm{J}/2$. Thus, we define region diameters to be $\simeq \lambda_{\mathrm{J}}$ as plotted in Fig.~\ref{fig.sizelumsigma}. We fit equation~(\ref{eq.sigr}) to the first panel in Fig.~\ref{fig.sizelumsigma} and obtain a normalisation for the gas surface density of $\Sigma_\mathrm{g} = 600$ \Msun~pc$^{-2}$.

\begin{figure}
\includegraphics[scale=0.52, trim = 11mm 0mm 6mm 2mm, clip]{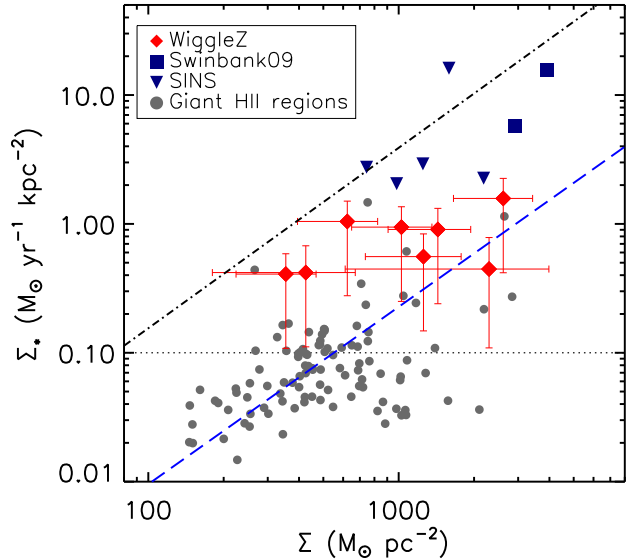}
\caption{Star formation surface density ($\Sigma_*$) vs. gas mass surface density ($\Sigma_\mathrm{g}$; derived from equation \ref{eq.sigr}) for the same data described in Fig.~\ref{fig.sizelumsigma} with all giant \HII regions now shown by gray points. The black dot-dash line is the Kennicutt-Schmidt law, $\Sigma_* \propto \Sigma_\mathrm{g}^{1.4}$ from \citet{2007ApJ...671..333K}. The average gas surface density for the WiggleZ clumps, assuming the KS law, derived from the measured $\Sigma_*$ is $\sim500$ \Msun~pc$^{-2}$ in comparison to what is derived from the Jeans equations for a pure gas disc of $\Sigma_\mathrm{g} \sim1000$ \Msun~pc$^{-2}$. The blue dashed line is the least squares fit to the data given by $\Sigma_*[\mathrm{\Msun~yr}^{-1}~\mathrm{kpc}^{-2}]=2\times10^{-5}~\Sigma_\mathrm{g}[\mathrm{\Msun~pc}^{-2}]^{1.37\pm0.14}$. The horizontal line represents the threshold of star formation required to host stellar winds \citep{2000ApJS..129..493H}. The star formation surface density and KS law are plotted for a Baldry-Glazebrook IMF.}
\label{fig.ks}
\end{figure}

We use the correlation between size and dispersion defined in equation (\ref{eq.sigr}) to estimate the gas surface density and test the Kennicutt-Schmidt law for individual star-forming regions. We assume that at the time of collapse, star-forming regions should have comparable surface densities and velocity dispersions to the disc in which they formed, and that the mass of the star-forming regions is dominated by gas ($\Sigma_\mathrm{g} = \sqrt{2}\sigma_\mathrm{cl}^2/\mathrm{G}r_\mathrm{cl}$). This is only an approximation as the regions are not purely molecular. Therefore, it is expected that the gas surface density determined by equation (\ref{eq.sigr}) should always be higher than the molecular gas surface density as estimated from the KS law. In Fig.~\ref{fig.ks} the star formation surface density is plotted against the derived gas surface density for the low and high-redshift regions. The KS law \citep{2007ApJ...671..333K} is overplotted by the black dot-dash line corrected to a BG03 IMF. The derived gas surface density for all regions is indeed greater than expected from the KS law, but is otherwise a reasonable approximation to the true gas surface density. The linear least squares fit to the data gives a slope of $1.38\pm0.14$ as shown by the blue dashed line and correlation coefficient of 0.66, consistent with the slope of 1.4 predicted by the KS law. We note that the fit degrades at high redshift. This could be to a selection effect given that we can only observe regions with the highest star formation surface densities. 

Using this method we obtain upper limits on the gas surface densities of order $\sim500-3000$ \Msun~pc$^{-2}$ for WiggleZ clumps, which overlap with the star formation surface densities for giant \HII regions ($\sim100-3000$ \Msun~pc$^{-2}$), and are comparable to the mean gas surface density for clumps at $z\sim2$ \citep{2010MNRAS.404.2151C,2011ApJ...733..101G}.  To emphasise the difference between the local and high-redshift environments in these instances, clumps have been compared to local giant molecular clouds (GMCs) which have typical surface densities of $\sim60-170$ \Msun~pc$^{-2}$ \citep{2007prpl.conf...81B,2008ApJ...686..948B,2009ApJ...699.1092H}. However, we argue that clumps are more comparable observationally and physically to \HII regions than GMCs as they are observed from their OB stars, which have already formed out of molecular gas.
 
The size-velocity dispersion relation shows the most scatter in Fig.~\ref{fig.sizelumsigma} with correlation coefficient of 0.72. Is this scatter due to measurement errors or an expected spread in region surface density in accordance with the KS law? The scatter in velocity dispersion ranges from $\sim20-90$ km s$^{-1}$ while the average error on the measured velocity dispersion of the WiggleZ clumps is $\sim20$ km s$^{-1}$. This scaling is expected to show scatter due to measurement errors discussed in Section 3:  size is the most uncertain parameter to measure, and the contribution of galactic winds in and around star-forming regions may result in added scatter in the velocity dispersion measurement. However, if regions are expected to follow the KS law (as Fig.~\ref{fig.ks} tentatively suggests) then from the three order-of magnitude spread measured in $\Sigma_*$ a range of $\sim4000~\Msun \mathrm{pc}^{-2}$ in $\Sigma_\mathrm{g}$ would be expected. This range translates into a spread in velocity dispersion of $\sim60$ km s$^{-1}$ at a given radius, approximately equivalent to the average scatter in Fig.~\ref{fig.sizelumsigma} and greater than the typical errors on velocity dispersion of the WiggleZ sample. Thus, if the KS law holds, then the expected variations in the gas surface density can explain the scattered relation between size and dispersion.

The same scaling between size and velocity dispersion, $\sigma \propto r^{1/2}$, can be obtained for virialised objects from the equation of dynamical mass ($M_\mathrm{dyn}=5\sigma^2r_\mathrm{cl}/\mathrm{G}$) and assuming a constant surface density such that $\sigma^2 =\pi\mathrm{G}\Sigma r_\mathrm{cl}/5$. When this relation is fit to the data in Fig.~\ref{fig.sizelumsigma} the best-fit average mass density for all regions, $\Sigma = 600$ \Msun~pc$^{-2}$, is equivalent to the surface density derived using the Jeans scaling.  From the scatter in Fig.~\ref{fig.sizelumsigma} the mass densities for all regions ranges from $100-5000$ \Msun~pc$^{-2}$ similar to what is seen in Fig.~\ref{fig.ks} and discussed above.  If regions have virialised then their mass will be better represented by the dynamical mass, which follows the same scaling as the Jeans mass ($M\propto \sigma^2r/\mathrm{G}$) determined in Section 4. However, the virial scaling yields masses a factor of $3\times$ greater than their Jeans mass. When the dynamical clump masses are added within their respective discs they approximate $\sim50-90$\% of the total stellar mass, compared to the Jeans masses which amount to $\sim20-30$\% of the total stellar masses of the discs. The Jeans masses are more comparable to the expected masses from individual clumps, and to the total mass contribution of clumps to the whole system, predicted by simulations \citep{2009ApJ...703..785D,Krumholz:2010fk,2011arXiv1106.5587C}. Thus, it seems unlikely that all the observed clumps have virialised and constitute such a large mass fraction of their discs.

We note that the high-redshift clumps, from the WiggleZ and SINS datasets, on average have higher velocity dispersions than predicted by the best-fit relation described by equation~(6). If we assume that the clumps from both samples are contaminated by galactic scale winds, as suggested here and in \cite{2011ApJ...733..101G}, with an average contribution comparable to the result of our stacking analysis in Section~\ref{sec.winds}, then the high-redshift dispersions are reduced by a factor of $\sim0.85$ making them more consistent with the scaling relation of equation~(6).

\begin{figure*}
\vspace*{0.2 cm}\includegraphics[scale=0.88]{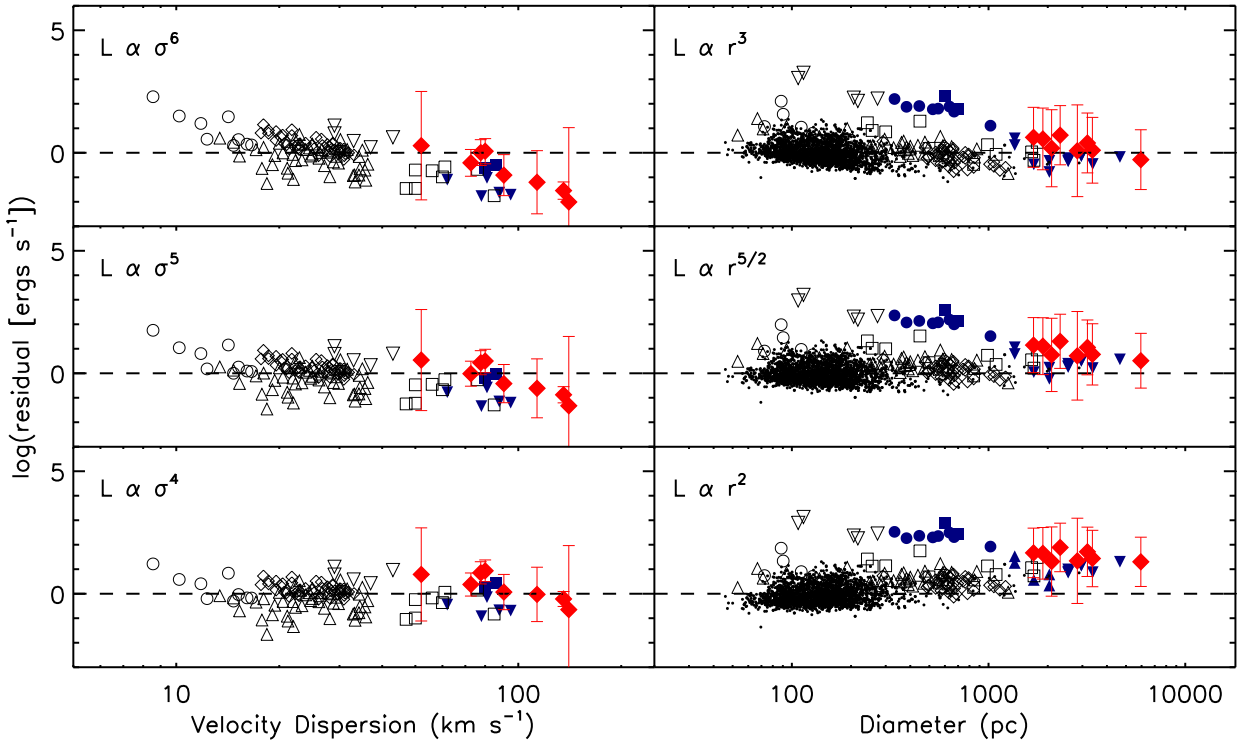}
\vspace{1.5 cm}
\caption{Residual luminosity vs. velocity dispersion and diameter from linear least squares fits assuming the scaling relations between luminosity and velocity dispersion and luminosity and size denoted in the upper left corner of each panel.  Symbols and colours are the same as in Fig.~\ref{fig.sizelumsigma}. The relations are derived for \textit{(top)} ionisation bound \stromgren~spheres that formed from Jeans collapse in an isothermal disc, \textit{(middle)} GMCs, and \textit{(bottom)} virialised regions. The $L-\sigma$ relation is best approximated by $L\propto\sigma^4$ and the $L-r$ relation is best approximated by $L\propto r^3$. These scaling relations are discussed in detail in Sections 5.2 and 5.3. The most deviant points in the right panels, offset by two orders of magnitude in luminosity, are from locally LIRG merger of NGC4038 and NGC4039 (open inverted triangles), lensed sub-L$_{*}$ galaxies at $z\sim2$ (blue closed circles), and an L$_{*}$ galaxy at $z\sim5$ (blue closed squares). The offset could be due to a difference in how regions are measured, as both high-redshift studies used the isophote method rather than profile fitting and the determination of sizes in the low-redshift sample is unclear.}\vspace*{0.3cm}
\label{fig.residuals}
\end{figure*}

\subsection{Luminosity vs. Size}
The scaling of \halpha luminosity with size of \HII regions is well understood if size is represented by the radius of a \stromgren~sphere, $r_\mathrm{s}$, the boundary between ionised and neutral hydrogen. The luminosity of an ionisation-bound region scales as L$_{\mathrm{H}\alpha}\propto r^3$, or more specifically
\begin{eqnarray}
L_{\mathrm{H}\alpha} = \frac{4\pi}{3}~\frac{\mathrm{N_H}^2\alpha_\mathrm{B}hc}{\lambda_{\mathrm{H}\alpha}}~r_\mathrm{s}^3,
\end{eqnarray}
where N$_\mathrm{H}$ is the ionised hydrogen number density and $\alpha_\mathrm{B}$ is the Case-B recombination coefficient assuming $T_\mathrm{HII}=10~000$ K \citep{1989agna.book.....O}. We obtain a density of N$_\mathrm{H} \sim 500$ cm$^{-3}$ by normalising the scaling relation to the data and assuming $r_\mathrm{s}=r_\mathrm{cl}$. For comparison, N$_\mathrm{H}\sim 100$ cm$^{-3}$ in local ULIRGS \citep{2007A&A...472..421M}. The residual from subtracting a least squares fit, fixing $L\propto r^3$, from the data is shown in Fig.~\ref{fig.residuals}. However, this is an idealised relation which assumes direct observations of individual \stromgren~spheres at a constant number density and temperature. The empirical relation observed in our data yields a slightly shallower slope of $2.72\pm0.04$, which could result from a variety of factors that act alone or in aggregate. Below we investigate the physical and observational processes that could affect the slope of the relation as well as consider the scalings of GMCs.

If the hydrogen atoms within the cloud can recombine faster than they can be ionised then not all the ionised photons will be absorbed. In this case, the ionisation region is said to be density bound, as the observed luminosity is limited by the density of the cloud, and could explain deviations from the scaling relation such that regions are less luminous at a given radius.  A detailed discussion of this scenario can be found in \cite{2000AJ....119.2728B} and the references therein. In this study all regions with $L_{\mathrm{H}\alpha}>10^{38.6}$ ergs s$^{-1}$ are predicted to be density bound due to a constant surface density for GMCs from which they form. We note that all of the WiggleZ clumps and the majority of giant \HII regions studied here are more luminous than this threshold, which suggests that a shallower slope could be observed as a result.  An important consequence of regions being density bound is that escaping photons could be responsible for ionising the diffuse interstellar medium in the disc.

As discussed in Section 3.1.3, resolution can affect the size-luminosity scaling relation. Beam-smearing results in a steeper slope in size-luminosity space than expected for resolved regions \citep{2000A&A...361..913P,Swinbank:2009tw}. Therefore, if the \stromgren~sphere scaling is the correct theoretical prediction for the empirical relation of size-luminosity, then resolution is not the dominant effect on the slope and may counterbalance other processes that result in a shallower slope (e.g. when regions are density bound). Dust, metallicity, and magnetic fields are also likely to reduce the observed ionised radiation at a given size, however a detailed discussion of their effects is beyond the scope of this paper. 

If the more appropriate analogues to \halpha clumps are GMCs rather than \HII regions as suggested in the literature \citep{2010ApJ...709..191M,Krumholz:2010fk} then a different scaling is expected between luminosity and size. For this case, a molecular tracer of radiation is necessary. The scaling between CO luminosity and size for molecular clouds follows a shallower relation, than expected from \stromgren~spheres, of order L $\propto r^{5/2}$ \citep{1987ApJ...319..730S,1989ApJ...338..178E}. However, although this scaling provides a reasonable fit to the \HII region data shown in Fig~\ref{fig.residuals}, it assumes a CO luminosity and cannot be tested by our \halpha data. We note that a similar relationship, of order $L\propto r^3$, has been observed between lensed clumps at $z=2.3$, GMC cores, and young \HII regions for $L_{260}$, a tracer of molecular gas \citep{2010Natur.464..733S}. These results suggest that the underlying physics of the star-forming processes is similar from $z=2.3$ to the present-day Universe.

Some observations \citep{Swinbank:2009tw,Jones:2010uf} of high-redshift clumps show deviations from the empirical luminosity-size relation. In these results the clumps in both a lensed L$_*$ galaxy at $z\sim5$ and lensed sub-L$_*$ galaxies at $z\sim2$ are offset by two orders of magnitude in luminosity from local \HII regions (blue filled circles and squares in Fig.~\ref{fig.sizelumsigma} and Fig.~\ref{fig.residuals}). An interpretation of the observed offset is that these regions, as well as giant \HII regions, form due to a starburst mode or due to lower metallicities of high-redshift galaxies. In comparison, there is a half order of magnitude offset in luminosity of the WiggleZ and SINS clumps from the best-fit relation to all the regions. A fraction of the offset in the WiggleZ and SINs clumps is likely due to an inflation of the clump luminosities from large-scale winds. The analysis in Section~\ref{sec.winds} suggests that winds can contribute up to 40\% of the luminosity of a region. If the luminosity of the WiggleZ and SINS clumps are corrected for this effect the remaining offset is less than half an order of magnitude from the best-fit relation. While a slight offset in luminosity exists for the WiggleZ clumps, it is small in comparison to the two order of magnitude offset found for the lensed clumps. The remaining offset could be a result of the factors discussed in this section. We find no significant luminosity offset between \HII regions and giant \HII regions at a given size. Furthermore, we include data from other low-metallicity regions with comparable values and find that they fall on the main relation. Thus, another basis is required to explain the deviation of the lensed regions. One possible explanation for the discrepancy is that the sizes of the regions from the two lensed studies are measured by the isophote method which could overestimate the luminosity for a given region by including surrounding flux of the disc.  

\subsection{Luminosity vs. Velocity Dispersion}
Assuming from our discussion above that \HII regions and clumps are idealised \stromgren~spheres ($L\propto r^3$), which form out of Jeans collapse in an isothermal disc ($r\propto \sigma^2$), then it is expected for luminosity to scale with dispersion as $L\propto \sigma^6$. This scaling can also be derived from the gas accretion rate as a function of virial velocity and the energy deposition rate due to accretion \citep{2008MNRAS.383..119D,2011A&A...530L...6L}. However, as outlined in the previous section and shown in Fig.~\ref{fig.residuals}, the regions observed are not idealised \stromgren~spheres and therefore a shallower slope is expected for the luminosity-velocity dispersion relation. An un-weighted fit to the data reveals $\log(L) = (4.18\pm 0.21)\times \log(\sigma) +  (33.61\pm 0.31)$.

We re-address the argument that the regions are virialised, given that the non-weighted fit most closely approximates $L \propto \sigma^4$ as the best description of the data. This scaling is expected from the virial theorem assuming a constant mass to light ratio ($M/L$) and constant surface brightness \citep{1976ApJ...204..668F}.  However, as $M/L$ is not constant (as shown in Fig~\ref{fig.mdyn}) deviations from the predicted scaling relations are expected and $L \propto \sigma^4$ is no longer an appropriate test of virialisation. Furthermore, the same scaling, assuming $M/L$ is constant, can be derived from the Jeans equations such that $M_\mathrm{J} \propto\sigma^4/\mathrm{G}^2\Sigma_\mathrm{g}$. 

\begin{figure}
\includegraphics[scale=0.52, trim = 11mm 0mm 6mm 2mm, clip]{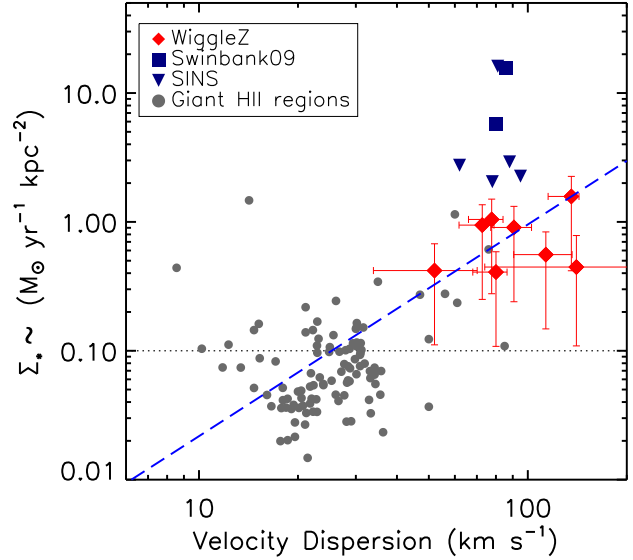}
\caption{Star formation surface density vs. velocity dispersion for the same data described in Fig.~\ref{fig.sizelumsigma} with all giant \HII regions shown by gray points. The blue dashed line is the least squares best fit to the data, given by $\sigma[\mathrm{km~s}^{-1}] = (100\pm 70 \times \Sigma_*[\mathrm{\Msun~yr}^{-1}~\mathrm{kpc}^{-2}])^{0.6\pm0.1}$. Although this relationship is more scattered than the $L_{\mathrm{H}\alpha}-\sigma$ relationship shown in Fig.~\ref{fig.sizelumsigma} a correlation is seen, which could be a result of mechanical energy released by the star formation ($\sigma = (\epsilon\Sigma_*)^{0.5}$; \citealt{2009ApJ...699.1660L}) or the expected star formation density in a marginally stable galactic disc of constant circular velocity assuming the Kennicutt-Schmidt law ($\sigma \propto \pi\mathrm{G}\Sigma_*^{0.7}/\kappa$; \citealt{Krumholz:2010fk}). The star formation surface density is plotted for a Baldry-Glazebrook IMF. }
\label{fig.sfsd}
\end{figure}

Finally, a relationship between luminosity and velocity dispersion of galaxies and the interstellar medium has recently been observed locally \citep{Dib:2006fk,Green:2010fk} and at high-redshift \citep{2009ApJ...699.1660L,Green:2010fk,2011ApJ...733..101G}. These observations have led the authors to conclude that star formation drives the large observed line-widths. \cite{2009ApJ...699.1660L} suggest that the velocity dispersion is due to mechanical energy released by the star formation in the form $\sigma = (\epsilon\Sigma_*)^{0.5}$ where $\epsilon$ is the efficiency of star formation and both $\sigma$ and $\Sigma_*$ can be global or local properties. The relationship between star formation surface density and velocity dispersion for the star-forming regions studied here is shown in Fig.~\ref{fig.sfsd} with the best linear least squares fit of $\sigma = (100\pm 70 \times \Sigma_*)^{0.6\pm0.1}$ represented by the blue dashed line. The fit is consistent with the results of \cite{2009ApJ...699.1660L}. However, the fit is also consistent with the expected star formation density in a marginally stable galactic disc of constant circular velocity assuming the KS law on galactic and sub-galactic scales, as noted by \cite{Krumholz:2010fk}. In this scenario, placing the KS law in equation (\ref{eq.toomre1}), assuming $Q=1$, yields $\sigma \propto \pi\mathrm{G}\Sigma_*^{0.7}/\kappa$. This interpretation is more consistent with our results in Section 5.1 and assumes the KS law holds for star-forming regions (Fig.~\ref{fig.ks}). However, we note that the $\sigma-\Sigma_*$ relation  shows a much greater scatter ($R^2$=0.57) than the relation between luminosity and velocity dispersion ($R^2$=0.87).

In conclusion, we find continuous scaling relations at high and low redshift for star-forming regions between \halpha size, luminosity, and velocity dispersion. When the three observables are combined by comparing \halpha luminosity with the Jeans mass ($M \propto \sigma^2 r_\mathrm{cl}/\mathrm{G}$) the scatter is significantly reduced ($L-\sigma: R^2=0.87$, $L-r: R^2=0.82$, and $L-M_\mathrm{J}: R^2=0.9)$. Some of the scatter observed in the size-velocity dispersion relation may be a natural consequence of the Kennicutt-Schmidt law such that $\sigma^2 \propto \Sigma_\mathrm{g}$. The data supports the theoretical arguments that star-forming regions form out of Jeans collapse in an isothermal disc with their luminosities approximated by \stromgren~spheres, with clumps accounting for 20-30\% of the stellar mass of the discs. These results indicate that star-forming regions at high-redshift form through similar processes as star-forming regions locally.

\section{Clump stability and survival}
It may be predicted as a consequence of the changes in global environment that star-forming regions at high redshift should not resemble star-forming regions locally. The high accretion rates predicted to drive up turbulence in the early universe are expected to dissipate on cosmological timescales for the most massive systems and contribute minimally at the present epoch \citep{2006MNRAS.368....2D}. Yet despite the increased sizes and dispersions, the regions presented here ($z\sim1.3$) have overlapping star formation surface densities and mass densities with giant \HII regions in the local universe and form continuous scaling relations with local star-formation regions. This indicates that while the turbulence of discs may have important implications for the size and luminosity of regions which form within them, the same processes govern their formation from high redshift to the current epoch.

Given the supersonic line-widths observed in the WiggleZ clumps and local \HII regions, star formation is expected to occur rapidly, depleting the available gas for star formation within a dynamical time. For the gas to remain turbulent as is observed, a mechanism is required to stir up the ISM \citep{1998ApJ...508L..99S,1998PhRvL..80.2754M}. Yet, there is no consensus in the field as to what generates gas velocity dispersions across these different epochs. The possible drivers of the turbulence at low-redshift are gravitational support \citep{1981MNRAS.195..839T}, hydrodynamical flows in the form of stellar winds \citep{1983ApJ...274..141G,1988A&A...201..199A,1993ApJ...418..767T}, champagne flows \citep{1983ApJ...274..141G}, supernova feedback \citep{2004RvMP...76..125M,2009ApJ...704..137J}, and bow shocks \citep{1993ApJ...418..767T}. These general scenarios have also been put forward to account for what maintains and drives up the turbulence at high redshift as well \citep{2004ApJ...611...20I,2007ApJ...670..237B,2006ApJ...645.1062F,2006Natur.442..786G,2009ApJ...703..785D,Krumholz:2010fk,2010ApJ...719.1230A,2010ApJ...712..294E,2010ApJ...719..229G} in addition to gas accretion \citep{2006ApJ...645.1062F,2006Natur.442..786G,2010ApJ...712..294E,2010ApJ...719..229G} and star formation \citep{2009ApJ...699.1660L,Green:2010fk}. In this section, the available data and models are compared where possible to explore the implications of the previous sections in the context of clump formation and lifetimes.


Models predict that the dominant feedback mechanism at high redshift is likely radiation pressure which can drive up the turbulent motions in the ISM \citep{Krumholz:2010fk,2010ApJ...709..191M}. Radiation feedback can provide pressure support to the clumps as well as drive large stellar winds.  However, if strong enough, radiation feedback can act to disrupt the clumps before virialising and migrating to the centre to form a stellar bulge \citep{2010ApJ...709..191M,Genel:2010vn}. Injection of energy into the interstellar medium by radiation pressure could be confirmed by detecting massive radiation-driven outflows from clumps \citep{Krumholz:2010fk}. In Section~\ref{sec.winds} an underlying broad component is detected, which contributes $\sim40$\% to the total emission line flux from the clumps. The broad \halpha component represents feedback due to thermal pressure from ionised gas and shocked stellar winds. These results, in combination with similar findings from $z\sim2$ clumps \citep{2011ApJ...733..101G}, provide evidence that winds from clumps could inject energy into the system, which would in turn drive up the turbulent motions in the ISM in accordance with the theoretical models.


If the clumps do survive feedback and clump-clump interactions then they may coalesce to form a bulge with high velocity dispersion $\sim600$ Myrs into the clumpy phase \citep{Elmegreen:2008fk}. From the kinematic sample of 13 WiggleZ galaxies studied in Paper I, five galaxies show a single resolved region of emission. Two of these galaxies show rotation, and could be interpreted as the final coalescence stage predicted by the models that include clump migration. These galaxies have average stellar masses of $3\times10^{10}$ \Msun~and are still forming stars at a rate of 30 \sfrunits~with velocity dispersions of 100 km s$^{-1}$. In the kinematic sample 3 galaxies exhibit multiple clumps, 5 are confused with extended regions of emission and 5 show single resolved regions of emission. The 10 galaxies not studied in detail in this paper may also be `clumpy' galaxies but due to projection or resolution effects individual regions are not distinguishable. However, the clumpy phase of these discs is expected to be short and thus we do not expect all galaxies observed to be in the early clumpy stage.

A different interpretation of the high-redshift data suggests that the turbulent motions arise from star formation \citep{2009ApJ...699.1660L}. This analysis, based on the correlation between disc velocity dispersions and \halpha luminosity, is supported locally by a study of luminous galaxies at $z\sim0.1$ \citep{Green:2010fk} and in models \citep{2010ApJ...712..294E}. We also find a tight scaling relation between luminosity and velocity dispersion (Fig.~\ref{fig.sizelumsigma}) and a noisier correlation between star formation surface density and velocity dispersion (Fig.~\ref{fig.sfsd}). However, these two parameters may be linked by the high dispersions within a Toomre unstable disc which drive up the star formation via the KS law, such that $\sigma \propto \Sigma_*^{1.4}$ \citep{Krumholz:2010fk}, as discussed in the previous Section. Unfortunately these arguments raise the problem of which is the cause and effect, high turbulence or high star formation rates?  We note that the mechanisms we have suggested to generate turbulence could also be interpreted as driving up star formation.

\cite{2011arXiv1106.5587C} argue that because turbulent motion can dissipate over a dynamical time, $\sim10\times$ shorter than clump lifetimes, then shortly after forming clumps must be supported against total free-fall collapse by internal rotation. Rotation also provides some support to clumps in the models of \cite{2009ApJ...703..785D}, \cite{2009MNRAS.397L..64A}, and \cite{2010ApJ...719.1230A}. In simulations, clumps have rotation velocities $\sim120$ km s$^{-1}$ and velocity dispersions comparable to disc velocity dispersions \citep{2011arXiv1106.5587C}. However, in this simulation the rotational signature of clumps is found to be reduced by a factor of 7 in observations for beam-smearing of 0.1 arcsec, comparable to the data from current IFS studies. This implies that the expected observed velocity gradients of clumps, $V_\mathrm{rot}\sim10-40$ km s$^{-1}$, are not measurable above the errors with current instrumentation. The clumps extracted from discs in these simulations have similar properties to the observed star-forming regions, as shown in the first panel of Fig.~\ref{fig.sizelumsigma}. The simulated and observed regions overlap in size and dispersion with a similar scatted in velocity dispersion at a given radius.

We test this model using the prescription of \cite{2011arXiv1106.5587C} which requires that clumps are rotationally supported when $V_\mathrm{rot}^2 / V_\mathrm{circ}^2 >0.5$ or $V_\mathrm{rot}/\sigma_\mathrm{1D} >\sqrt{2}$. 
We estimate for the WiggleZ clumps $V_\mathrm{rot}/\sigma_\mathrm{1D}$ in which $\sigma_\mathrm{1D}\sim\sigma_\mathrm{net}$ and $V_\mathrm{rot} \sim v_\mathrm{shear} \times 7$ (correcting for beam-smearing assuming by a factor of 7) where $v_\mathrm{shear}$ is the velocity shear across the clumps as measure from the average of the maximum and minimum velocity residuals within the clump radius ($\sim 20-80$ km s$^{-1}$ for the WiggleZ clumps). Using this prescription we find that all regions could be rotationally supported. However, these conclusions are uncertain given our measurement errors. Observed clumps at $z\sim2$ show tentative evidence for rotational support for a fraction of the sample, with velocity shear measured from residual \halpha velocity maps after model discs have been subtracted \citep{2011ApJ...733..101G}.  Together, these results suggest rotation is not likely to be the primary support mechanism for the observed clumps. If clumps survive long enough to virialise then they could become rotationally supported. A more suitable test would be to observe the kinematics of more evolved clumps, seen in broadband imaging, which may have had time to virialise.  

In conclusion, there are many processes internal and external to galaxies that are able to inject energy into the ISM to maintain the high observed velocity dispersions. The WiggleZ clumps have consistent low metallicity across the discs they are embedded in, indicating that they are young with common evolution histories. They do not show evidence of rotation suggesting that they have not virialised. However, this interpretation is limited by our data due to the effects of beam-smearing. Higher S/N and higher resolution data of many more clumps are needed to provide more rigourous tests of the existing theories of clump evolution at high redshift.

\vspace*{-0.4cm}\section{Conclusions}

We present 8 new clumps embedded within unstable disc galaxies at $z\sim1.3$ from the kinematic sub-sample of the WiggleZ Dark Energy Survey \citep{2011MNRAS.417.2601W}. The clumps have the following properties:
\begin{itemize}
\item Average clump size of $r_\mathrm{core}\sim1.5$ kpc. 
\item Average clump velocity dispersions of $\sigma\sim90$ km s$^{-1}$, higher than all other clump samples at $z>1$.
\item Average Jeans masses of $4.2\times10^{9}$ \Msun, in total accounting for 20-30\% of the stellar mass of the discs.
\item Estimated gas mass surface densities in the range $\Sigma_\mathrm{g}=300-2000$ \Msun~pc$^{-2}$.
\item Clump star formation surface densities of $\Sigma_*=0.4-2$ \Msun~yr$^{-1}$ kpc$^{-2}$, with total clump star formation accounting for $30-40$\% of star formation in the discs (BG03 IMF). 
\item In the stacked spectrum of all clumps we detect a broad \halpha component of FWHM $\sim$ 490 km s$^{-1}$ which contributes 40\% to the overall clump flux and is likely due to the presence of large-scale stellar winds driven by intense star formation. This conclusion is supported by the condition that all clumps are above the threshold required to drive galactic scale winds, $\Sigma_*>0.1$ \Msun~ yr$^{-1}$ kpc$^{-2}$. 
\end{itemize}

Clumps, \HII regions, and giant \HII regions are found to follow the same scaling relations for \halpha size, velocity dispersion, and luminosity, indicating that they may be scaled-up versions of local \HII regions. Our best-fitting relations are $\log(\sigma) =  (0.42\pm 0.03)\times\log(d) + ( 0.33\pm 0.09)$, $\log(L_{\mathrm{H}\alpha})=  (2.72\pm 0.04)\times\log(d) + (31.99\pm 0.08)$, and  $\log(L_{\mathrm{H}\alpha})=  (4.18\pm 0.21)\times\log(\sigma) + (33.61\pm 0.31)$. We show that these relationships hold over four orders of magnitude in mass ($10^6-10^{10}$ \Msun) from \HII regions to giant clumps at high redshift. The size-velocity dispersion relation is consistent with Jeans collapse in an isothermal disc with a range in star formation surface density of $\sim4000$ \Msun~pc$^{-2}$. The regions closely approximate the expected relation for idealised \stromgren~spheres although they may not be completely ionisation bound as a slightly shallower relation is observed. This could be due to a range of factors including resolution, cloud density, dust, metallicity, and magnetic fields. 

The relationship between luminosity and velocity dispersion is not as well understood. We find an empirical relation of $L\propto \sigma^4$, which deviates from the predicted relation $L\propto \sigma^6$ assuming that star-forming regions form under Jeans collapse and their luminosities are driven by the \stromgren~sphere physics. A possible explanation for the empirical relation is that the luminosity closely traces the mass, such that $M\propto\sigma^4/\mathrm{G}^2\Sigma$, a scaling obtained for both the Jeans and virial masses. We find that the Jeans mass yields a mass in better agreement with models than the dynamical mass and thus is the more likely driver of the observed correlation.

The data is consistent with high surface densities as a result of higher velocity dispersions in a marginally unstable disc assuming a Kennicutt-Schmidt law \citep{Krumholz:2010fk}. This is a reasonable approximation given that we find tentative evidence that star-forming regions follow the Kennicutt-Schmidt law. If this star-formation law holds then an expected range in the gas surface density would explain the high scatter in the $\sigma-r$ relation and would predict gas surface densities in agreement with estimates from the Jeans equations.

We do not find a luminosity offset between our clumps and local \HII regions as reported for other high-redshift studies \citep{Swinbank:2009tw,Jones:2010uf}. We extract measurements from the literature for giant \HII regions in isolated spirals, (U)LIRGS, irregulars, and dwarf irregulars. After correcting historical data for updated distances and cosmologies we find that \HII regions and giant \HII regions follow a single size-luminosity relationship. A few outliers persist locally from the LIRG merger of NGC4038 and NGC4039 \citep{2006A&A...445..471B} and sub-L$_{*}$ galaxies at $z\sim2$ \citep{Jones:2010uf} and an L$_{*}$ galaxy at $z\sim5$ \citep{Swinbank:2009tw}. This offset could be due to a difference in how regions are measured, as both high-redshift studies used the isophote method rather than profile fitting and the determination of sizes in the low-redshift sample is unclear.

We determine that the most consistent method to measure the size of an \HII region is from its core radius determined by fitting a Gaussian profile to the central brightest region of \halpha emission. Theoretically, this measurement is based on the formation of idealised fully ionised regions that are not contaminated along the line of sight with intervening gas and dust. Observationally, this measurement is insensitive to variations in the background, however it becomes difficult to measure in low signal-to-noise data.

The results presented here are in relative agreement with massive star-forming regions forming out of Toomre-unstable discs with high local velocity dispersion, with measured sizes and estimated masses in agreement with the model predictions (e.g. \citealt{1999ApJ...514...77N,2004A&A...413..547I,2004ApJ...611...20I,2007ApJ...670..237B,Elmegreen:2008fk}). The data supports models that employ radiation feedback in the form of large stellar winds driven by regions of enhanced star formation \citep{Krumholz:2010fk,Genel:2010vn}, however we cannot say on what time scale the winds would destroy the clumps. The dominant mechanism stabilising clumps from rapid collapse is unclear. Given the importance of high velocity dispersions in the formation of large massive star-forming regions and minimal signature of rotation, turbulence is likely an important contributor. For the discs to become initially unstable the injection of energy from cold accretion is the probable cause. However, different mechanisms may dominate throughout the lifetime of the discs, such as disc instabilities and feedback from young stars, both of which are observed here \citep{2010ApJ...712..294E}. There are many possible mechanisms that may produce local velocity dispersions $-$ inflow, winds, shocks, gravitational energy, star formation $-$ further research is needed to understand the interplay of their significance throughout the lifetimes of high-redshift discs. 

\clearpage

\section*{Acknowledgments}
We thank the referee for providing valuable comments. EW thanks Max Malacari and Nadine Bachmann for their valuable help on measuring clump sizes and Mark Swinbank and Rachel Livermore for useful discussions.

Some of the data presented herein were obtained at the W.M. Keck Observatory, which is operated as a scientific partnership among the California Institute of Technology, the University of California and the National Aeronautics and Space Administration. The Observatory was made possible by the generous financial support of the W.M. Keck Foundation. The authors wish to recognise and acknowledge the very significant cultural role and reverence that the summit of Mauna Kea has always had within the indigenous Hawaiian community.  We are most fortunate to have the opportunity to conduct observations from this mountain.

\bibliographystyle{apj}

\begin{thebibliography}{98}
\expandafter\ifx\csname natexlab\endcsname\relax\def\natexlab#1{#1}\fi

\bibitem[{{Agertz} {et~al.}(2009){Agertz}, {Teyssier}, \&
  {Moore}}]{2009MNRAS.397L..64A}
{Agertz}, O., {Teyssier}, R., \& {Moore}, B. 2009, \mnras, 397, L64

\bibitem[{{Alexander} {et~al.}(2010){Alexander}, {Swinbank}, {Smail},
  {McDermid}, \& {Nesvadba}}]{Alexander:2010bd}
{Alexander}, D.~M., {Swinbank}, A.~M., {Smail}, I., {McDermid}, R., \&
  {Nesvadba}, N.~P.~H. 2010, \mnras, 402, 2211

\bibitem[{{Arsenault} \& {Roy}(1988)}]{1988A&A...201..199A}
{Arsenault}, R., \& {Roy}, J. 1988, \aap, 201, 199

\bibitem[{{Aumer} {et~al.}(2010){Aumer}, {Burkert}, {Johansson}, \&
  {Genzel}}]{2010ApJ...719.1230A}
{Aumer}, M., {Burkert}, A., {Johansson}, P.~H., \& {Genzel}, R. 2010, \apj,
  719, 1230

\bibitem[{{Baldry} \& {Glazebrook}(2003)}]{2003ApJ...593..258B}
{Baldry}, I.~K., \& {Glazebrook}, K. 2003, \apj, 593, 258

\bibitem[{{Bastian} {et~al.}(2006){Bastian}, {Emsellem}, {Kissler-Patig}, \&
  {Maraston}}]{2006A&A...445..471B}
{Bastian}, N., {Emsellem}, E., {Kissler-Patig}, M., \& {Maraston}, C. 2006,
  \aap, 445, 471

\bibitem[{{Beckman} {et~al.}(2000){Beckman}, {Rozas}, {Zurita}, {Watson}, \&
  {Knapen}}]{2000AJ....119.2728B}
{Beckman}, J.~E., {Rozas}, M., {Zurita}, A., {Watson}, R.~A., \& {Knapen},
  J.~H. 2000, \aj, 119, 2728

\bibitem[{{Binney} \& {Tremaine}(2008)}]{2008gady.book.....B}
{Binney}, J., \& {Tremaine}, S. 2008, {Galactic Dynamics: Second Edition}, ed.
  {Binney, J.~\& Tremaine, S.} (Princeton University Press)

\bibitem[{{Blitz} {et~al.}(2007){Blitz}, {Fukui}, {Kawamura}, {Leroy},
  {Mizuno}, \& {Rosolowsky}}]{2007prpl.conf...81B}
{Blitz}, L., {Fukui}, Y., {Kawamura}, A., {Leroy}, A., {Mizuno}, N., \&
  {Rosolowsky}, E. 2007, Protostars and Planets V, 81

\bibitem[{{Bolatto} {et~al.}(2008){Bolatto}, {Leroy}, {Rosolowsky}, {Walter},
  \& {Blitz}}]{2008ApJ...686..948B}
{Bolatto}, A.~D., {Leroy}, A.~K., {Rosolowsky}, E., {Walter}, F., \& {Blitz},
  L. 2008, \apj, 686, 948

\bibitem[{{Bournaud} {et~al.}(2008){Bournaud}, {Daddi}, {Elmegreen},
  {Elmegreen}, {Nesvadba}, {Vanzella}, {Di Matteo}, {Le Tiran}, {Lehnert}, \&
  {Elbaz}}]{2008A&A...486..741B}
{Bournaud}, F., {et~al.} 2008, \aap, 486, 741

\bibitem[{Bournaud {et~al.}(2011)Bournaud, Dekel, Teyssier, Cacciato, Daddi,
  Juneau, \& Shankar}]{Bournaud:2011fk}
Bournaud, F., Dekel, A., Teyssier, R., Cacciato, M., Daddi, E., Juneau, S., \&
  Shankar, F. 2011

\bibitem[{{Bournaud} {et~al.}(2007){Bournaud}, {Elmegreen}, \&
  {Elmegreen}}]{2007ApJ...670..237B}
{Bournaud}, F., {Elmegreen}, B.~G., \& {Elmegreen}, D.~M. 2007, \apj, 670, 237

\bibitem[{{Burkert} {et~al.}(2010){Burkert}, {Genzel}, {Bouch{\'e}}, {Cresci},
  {Khochfar}, {Sommer-Larsen}, {Sternberg}, {Naab}, {F{\"o}rster Schreiber},
  {Tacconi}, {Shapiro}, {Hicks}, {Lutz}, {Davies}, {Buschkamp}, \&
  {Genel}}]{2010ApJ...725.2324B}
{Burkert}, A., {et~al.} 2010, \apj, 725, 2324

\bibitem[{{Ceverino} {et~al.}(2010){Ceverino}, {Dekel}, \&
  {Bournaud}}]{2010MNRAS.404.2151C}
{Ceverino}, D., {Dekel}, A., \& {Bournaud}, F. 2010, \mnras, 404, 2151

\bibitem[{{Ceverino} {et~al.}(2011){Ceverino}, {Dekel}, {Mandelker},
  {Bournaud}, {Burkert}, {Genzel}, \& {Primack}}]{2011arXiv1106.5587C}
{Ceverino}, D., {Dekel}, A., {Mandelker}, N., {Bournaud}, F., {Burkert}, A.,
  {Genzel}, R., \& {Primack}, J. 2011, ArXiv e-prints

\bibitem[{{Combes} {et~al.}(2011){Combes}, {Garc{\'{\i}}a-Burillo}, {Braine},
  {Schinnerer}, {Walter}, \& {Colina}}]{2011A&A...528A.124C}
{Combes}, F., {Garc{\'{\i}}a-Burillo}, S., {Braine}, J., {Schinnerer}, E.,
  {Walter}, F., \& {Colina}, L. 2011, \aap, 528, A124+

\bibitem[{{Cowie} {et~al.}(1995){Cowie}, {Hu}, \& {Songaila}}]{cowie:1995:10}
{Cowie}, L.~L., {Hu}, E.~M., \& {Songaila}, A. 1995, \aj, 110, 1576

\bibitem[{{Daddi} {et~al.}(2010){Daddi}, {Bournaud}, {Walter}, {Dannerbauer},
  {Carilli}, {Dickinson}, {Elbaz}, {Morrison}, {Riechers}, {Onodera}, {Salmi},
  {Krips}, \& {Stern}}]{2010ApJ...713..686D}
{Daddi}, E., {et~al.} 2010, \apj, 713, 686

\bibitem[{{Davies} {et~al.}(2011){Davies}, {F{\"o}rster Schreiber}, {Cresci},
  {Genzel}, {Bouch{\'e}}, {Burkert}, {Buschkamp}, {Genel}, {Hicks}, {Kurk},
  {Lutz}, {Newman}, {Shapiro}, {Sternberg}, {Tacconi}, \&
  {Wuyts}}]{2011ApJ...741...69D}
{Davies}, R., {et~al.} 2011, \apj, 741, 69

\bibitem[{{Dekel} \& {Birnboim}(2006)}]{2006MNRAS.368....2D}
{Dekel}, A., \& {Birnboim}, Y. 2006, \mnras, 368, 2

\bibitem[{{Dekel} \& {Birnboim}(2008)}]{2008MNRAS.383..119D}
---. 2008, \mnras, 383, 119

\bibitem[{{Dekel} {et~al.}(2009){Dekel}, {Sari}, \&
  {Ceverino}}]{2009ApJ...703..785D}
{Dekel}, A., {Sari}, R., \& {Ceverino}, D. 2009, \apj, 703, 785

\bibitem[{{Dib} {et~al.}(2006){Dib}, {Bell}, \& {Burkert}}]{Dib:2006fk}
{Dib}, S., {Bell}, E., \& {Burkert}, A. 2006, \apj, 638, 797

\bibitem[{{Dickinson}(2000)}]{2000bgfp.conf..257D}
{Dickinson}, M. 2000, in Building Galaxies; from the Primordial Universe to the
  Present, ed. {F.~Hammer, T.~X.~Thuan, V.~Cayatte, B.~Guiderdoni, \&
  J.~T.~Thanh Van }, 257--+

\bibitem[{{Drinkwater} {et~al.}(2010){Drinkwater}, {Jurek}, {Blake}, {Woods},
  {Pimbblet}, {Glazebrook}, {Sharp}, {Pracy}, {Brough}, {Colless}, {Couch},
  {Croom}, {Davis}, {Forbes}, {Forster}, {Gilbank}, {Gladders}, {Jelliffe},
  {Jones}, {Li}, {Madore}, {Martin}, {Poole}, {Small}, {Wisnioski}, {Wyder}, \&
  {Yee}}]{Drinkwater:2010bx}
{Drinkwater}, M.~J., {et~al.} 2010, \mnras, 401, 1429

\bibitem[{{Elmegreen}(1989)}]{1989ApJ...338..178E}
{Elmegreen}, B.~G. 1989, \apj, 338, 178

\bibitem[{{Elmegreen} {et~al.}(2008){Elmegreen}, {Bournaud}, \&
  {Elmegreen}}]{Elmegreen:2008fk}
{Elmegreen}, B.~G., {Bournaud}, F., \& {Elmegreen}, D.~M. 2008, \apj, 688, 67

\bibitem[{{Elmegreen} \& {Burkert}(2010)}]{2010ApJ...712..294E}
{Elmegreen}, B.~G., \& {Burkert}, A. 2010, \apj, 712, 294

\bibitem[{{Elmegreen} \& {Elmegreen}(2005)}]{2005ApJ...627..632E}
{Elmegreen}, B.~G., \& {Elmegreen}, D.~M. 2005, \apj, 627, 632

\bibitem[{{Elmegreen} {et~al.}(2009){Elmegreen}, {Elmegreen}, {Marcus},
  {Shahinyan}, {Yau}, \& {Petersen}}]{2009ApJ...701..306E}
{Elmegreen}, D.~M., {Elmegreen}, B.~G., {Marcus}, M.~T., {Shahinyan}, K.,
  {Yau}, A., \& {Petersen}, M. 2009, \apj, 701, 306

\bibitem[{{Faber} \& {Jackson}(1976)}]{1976ApJ...204..668F}
{Faber}, S.~M., \& {Jackson}, R.~E. 1976, \apj, 204, 668

\bibitem[{{Ferguson} {et~al.}(1996){Ferguson}, {Wyse}, {Gallagher}, \&
  {Hunter}}]{1996AJ....111.2265F}
{Ferguson}, A.~M.~N., {Wyse}, R.~F.~G., {Gallagher}, III, J.~S., \& {Hunter},
  D.~A. 1996, \aj, 111, 2265

\bibitem[{{F{\"o}rster Schreiber} {et~al.}(2009){F{\"o}rster Schreiber},
  {Genzel}, {Bouch{\'e}}, {Cresci}, {Davies}, {Buschkamp}, {Shapiro},
  {Tacconi}, {Hicks}, {Genel}, {Shapley}, {Erb}, {Steidel}, {Lutz},
  {Eisenhauer}, {Gillessen}, {Sternberg}, {Renzini}, {Cimatti}, {Daddi},
  {Kurk}, {Lilly}, {Kong}, {Lehnert}, {Nesvadba}, {Verma}, {McCracken},
  {Arimoto}, {Mignoli}, \& {Onodera}}]{2009ApJ...706.1364F}
{F{\"o}rster Schreiber}, N.~M., {et~al.} 2009, \apj, 706, 1364

\bibitem[{{F{\"o}rster Schreiber} {et~al.}(2006){F{\"o}rster Schreiber},
  {Genzel}, {Lehnert}, {Bouch{\'e}}, {Verma}, {Erb}, {Shapley}, {Steidel},
  {Davies}, {Lutz}, {Nesvadba}, {Tacconi}, {Eisenhauer}, {Abuter}, {Gilbert},
  {Gillessen}, \& {Sternberg}}]{2006ApJ...645.1062F}
---. 2006, \apj, 645, 1062

\bibitem[{{F{\"o}rster Schreiber} {et~al.}(2010){F{\"o}rster Schreiber},
  {Shapley}, {Erb}, {Genzel}, {Steidel}, {Bouch{\'e}}, {Cresci}, \&
  {Davies}}]{2010arXiv1011.1507F}
{F{\"o}rster Schreiber}, N.~M., {Shapley}, A.~E., {Erb}, D.~K., {Genzel}, R.,
  {Steidel}, C.~C., {Bouch{\'e}}, N., {Cresci}, G., \& {Davies}, R. 2010, ArXiv
  e-prints

\bibitem[{{F{\"o}rster Schreiber} {et~al.}(2011){F{\"o}rster Schreiber},
  {Shapley}, {Genzel}, {Bouch{\'e}}, {Cresci}, {Davies}, {Erb}, {Genel},
  {Lutz}, {Newman}, {Shapiro}, {Steidel}, {Sternberg}, \&
  {Tacconi}}]{2011ApJ...739...45F}
{F{\"o}rster Schreiber}, N.~M., {et~al.} 2011, \apj, 739, 45

\bibitem[{{Fuentes-Masip} {et~al.}(2000){Fuentes-Masip},
  {Mu{\~n}oz-Tu{\~n}{\'o}n}, {Casta{\~n}eda}, \&
  {Tenorio-Tagle}}]{2000AJ....120..752F}
{Fuentes-Masip}, O., {Mu{\~n}oz-Tu{\~n}{\'o}n}, C., {Casta{\~n}eda}, H.~O., \&
  {Tenorio-Tagle}, G. 2000, \aj, 120, 752

\bibitem[{{Gallagher} \& {Hunter}(1983)}]{1983ApJ...274..141G}
{Gallagher}, J.~S., \& {Hunter}, D.~A. 1983, \apj, 274, 141

\bibitem[{{Genel} {et~al.}(2010){Genel}, {Bouch{\'e}}, {Naab}, {Sternberg}, \&
  {Genzel}}]{2010ApJ...719..229G}
{Genel}, S., {Bouch{\'e}}, N., {Naab}, T., {Sternberg}, A., \& {Genzel}, R.
  2010, \apj, 719, 229

\bibitem[{Genel {et~al.}(2010)Genel, Naab, Genzel, Schreiber, Sternberg, Oser,
  Johansson, Dav{\'e}, Oppenheimer, \& Burkert}]{Genel:2010vn}
Genel, S., {et~al.} 2010

\bibitem[{{Genzel} {et~al.}(2011){Genzel}, {Newman}, {Jones}, {F{\"o}rster
  Schreiber}, {Shapiro}, {Genel}, {Lilly}, {Renzini}, {Tacconi}, {Bouch{\'e}},
  {Burkert}, {Cresci}, {Buschkamp}, {Carollo}, {Ceverino}, {Davies}, {Dekel},
  {Eisenhauer}, {Hicks}, {Kurk}, {Lutz}, {Mancini}, {Naab}, {Peng},
  {Sternberg}, {Vergani}, \& {Zamorani}}]{2011ApJ...733..101G}
{Genzel}, R., {et~al.} 2011, \apj, 733, 101

\bibitem[{{Genzel} {et~al.}(2006){Genzel}, {Tacconi}, {Eisenhauer},
  {F{\"o}rster Schreiber}, {Cimatti}, {Daddi}, {Bouch{\'e}}, {Davies},
  {Lehnert}, {Lutz}, {Nesvadba}, {Verma}, {Abuter}, {Shapiro}, {Sternberg},
  {Renzini}, {Kong}, {Arimoto}, \& {Mignoli}}]{2006Natur.442..786G}
---. 2006, \nat, 442, 786

\bibitem[{{Green} {et~al.}(2010){Green}, {Glazebrook}, {McGregor}, {Abraham},
  {Poole}, {Damjanov}, {McCarthy}, {Colless}, \& {Sharp}}]{Green:2010fk}
{Green}, A.~W., {et~al.} 2010, \nat, 467, 684

\bibitem[{{Guo} {et~al.}(2011){Guo}, {Giavalisco}, {Ferguson}, {Cassata}, \&
  {Koekemoer}}]{2011arXiv1110.3800G}
{Guo}, Y., {Giavalisco}, M., {Ferguson}, H.~C., {Cassata}, P., \& {Koekemoer},
  A.~M. 2011, ArXiv e-prints

\bibitem[{{Guti{\'e}rrez} {et~al.}(2011){Guti{\'e}rrez}, {Beckman}, \&
  {Buenrostro}}]{2011AJ....141..113G}
{Guti{\'e}rrez}, L., {Beckman}, J.~E., \& {Buenrostro}, V. 2011, \aj, 141, 113

\bibitem[{{Heckman} {et~al.}(1990){Heckman}, {Armus}, \&
  {Miley}}]{1990ApJS...74..833H}
{Heckman}, T.~M., {Armus}, L., \& {Miley}, G.~K. 1990, \apjs, 74, 833

\bibitem[{{Heckman} {et~al.}(2000){Heckman}, {Lehnert}, {Strickland}, \&
  {Armus}}]{2000ApJS..129..493H}
{Heckman}, T.~M., {Lehnert}, M.~D., {Strickland}, D.~K., \& {Armus}, L. 2000,
  \apjs, 129, 493

\bibitem[{{Heyer} {et~al.}(2009){Heyer}, {Krawczyk}, {Duval}, \&
  {Jackson}}]{2009ApJ...699.1092H}
{Heyer}, M., {Krawczyk}, C., {Duval}, J., \& {Jackson}, J.~M. 2009, \apj, 699,
  1092

\bibitem[{{Immeli} {et~al.}(2004{\natexlab{a}}){Immeli}, {Samland}, {Gerhard},
  \& {Westera}}]{2004A&A...413..547I}
{Immeli}, A., {Samland}, M., {Gerhard}, O., \& {Westera}, P.
  2004{\natexlab{a}}, \aap, 413, 547

\bibitem[{{Immeli} {et~al.}(2004{\natexlab{b}}){Immeli}, {Samland}, {Westera},
  \& {Gerhard}}]{2004ApJ...611...20I}
{Immeli}, A., {Samland}, M., {Westera}, P., \& {Gerhard}, O.
  2004{\natexlab{b}}, \apj, 611, 20

\bibitem[{{Jones} {et~al.}(2010){Jones}, {Swinbank}, {Ellis}, {Richard}, \&
  {Stark}}]{Jones:2010uf}
{Jones}, T.~A., {Swinbank}, A.~M., {Ellis}, R.~S., {Richard}, J., \& {Stark},
  D.~P. 2010, \mnras, 404, 1247

\bibitem[{{Joung} \& {Mac Low}(2006)}]{2006ApJ...653.1266J}
{Joung}, M.~K.~R., \& {Mac Low}, M.-M. 2006, \apj, 653, 1266

\bibitem[{{Joung} {et~al.}(2009){Joung}, {Mac Low}, \&
  {Bryan}}]{2009ApJ...704..137J}
{Joung}, M.~R., {Mac Low}, M.-M., \& {Bryan}, G.~L. 2009, \apj, 704, 137

\bibitem[{{Kennicutt}(1979)}]{1979ApJ...228..696K}
{Kennicutt}, Jr., R.~C. 1979, \apj, 228, 696

\bibitem[{{Kennicutt}(1998)}]{1998ARA&A..36..189K}
---. 1998, \araa, 36, 189

\bibitem[{{Kennicutt} {et~al.}(2003){Kennicutt}, {Armus}, {Bendo}, {Calzetti},
  {Dale}, {Draine}, {Engelbracht}, {Gordon}, {Grauer}, {Helou}, {Hollenbach},
  {Jarrett}, {Kewley}, {Leitherer}, {Li}, {Malhotra}, {Regan}, {Rieke},
  {Rieke}, {Roussel}, {Smith}, {Thornley}, \& {Walter}}]{2003PASP..115..928K}
{Kennicutt}, Jr., R.~C., {et~al.} 2003, \pasp, 115, 928

\bibitem[{{Kennicutt} {et~al.}(2007){Kennicutt}, {Calzetti}, {Walter}, {Helou},
  {Hollenbach}, {Armus}, {Bendo}, {Dale}, {Draine}, {Engelbracht}, {Gordon},
  {Prescott}, {Regan}, {Thornley}, {Bot}, {Brinks}, {de Blok}, {de Mello},
  {Meyer}, {Moustakas}, {Murphy}, {Sheth}, \& {Smith}}]{2007ApJ...671..333K}
---. 2007, \apj, 671, 333

\bibitem[{{Kere{\v s}} {et~al.}(2005){Kere{\v s}}, {Katz}, {Weinberg}, \&
  {Dav{\'e}}}]{2005MNRAS.363....2K}
{Kere{\v s}}, D., {Katz}, N., {Weinberg}, D.~H., \& {Dav{\'e}}, R. 2005,
  \mnras, 363, 2

\bibitem[{{Krumholz} \& {Burkert}(2010)}]{2010ApJ...724..895K}
{Krumholz}, M., \& {Burkert}, A. 2010, \apj, 724, 895

\bibitem[{{Krumholz} \& {Dekel}(2010)}]{Krumholz:2010fk}
{Krumholz}, M.~R., \& {Dekel}, A. 2010, \mnras, 406, 112

\bibitem[{{Larkin} {et~al.}(2006){Larkin}, {Barczys}, {Krabbe}, {Adkins},
  {Aliado}, {Amico}, {Brims}, {Campbell}, {Canfield}, {Gasaway}, {Honey},
  {Iserlohe}, {Johnson}, {Kress}, {Lafreniere}, {Magnone}, {Magnone},
  {McElwain}, {Moon}, {Quirrenbach}, {Skulason}, {Song}, {Spencer}, {Weiss}, \&
  {Wright}}]{2006NewAR..50..362L}
{Larkin}, J., {et~al.} 2006, \nar, 50, 362

\bibitem[{{Larson}(1981)}]{1981MNRAS.194..809L}
{Larson}, R.~B. 1981, \mnras, 194, 809

\bibitem[{{Law} {et~al.}(2007){Law}, {Steidel}, {Erb}, {Larkin}, {Pettini},
  {Shapley}, \& {Wright}}]{2007ApJ...669..929L}
{Law}, D.~R., {Steidel}, C.~C., {Erb}, D.~K., {Larkin}, J.~E., {Pettini}, M.,
  {Shapley}, A.~E., \& {Wright}, S.~A. 2007, \apj, 669, 929

\bibitem[{{Le Tiran} {et~al.}(2011){Le Tiran}, {Lehnert}, {Di Matteo},
  {Nesvadba}, \& {van Driel}}]{2011A&A...530L...6L}
{Le Tiran}, L., {Lehnert}, M.~D., {Di Matteo}, P., {Nesvadba}, N.~P.~H., \&
  {van Driel}, W. 2011, \aap, 530, L6

\bibitem[{{Lehnert} {et~al.}(2009){Lehnert}, {Nesvadba}, {Tiran}, {Di Matteo},
  {van Driel}, {Douglas}, {Chemin}, \& {Bournaud}}]{2009ApJ...699.1660L}
{Lehnert}, M.~D., {Nesvadba}, N.~P.~H., {Tiran}, L.~L., {Di Matteo}, P., {van
  Driel}, W., {Douglas}, L.~S., {Chemin}, L., \& {Bournaud}, F. 2009, \apj,
  699, 1660

\bibitem[{{Mac Low} \& {Klessen}(2004)}]{2004RvMP...76..125M}
{Mac Low}, M.-M., \& {Klessen}, R.~S. 2004, Reviews of Modern Physics, 76, 125

\bibitem[{{Mac Low} {et~al.}(1998){Mac Low}, {Klessen}, {Burkert}, \&
  {Smith}}]{1998PhRvL..80.2754M}
{Mac Low}, M.-M., {Klessen}, R.~S., {Burkert}, A., \& {Smith}, M.~D. 1998,
  Physical Review Letters, 80, 2754

\bibitem[{{Melnick} {et~al.}(1987){Melnick}, {Moles}, {Terlevich}, \&
  {Garcia-Pelayo}}]{1987MNRAS.226..849M}
{Melnick}, J., {Moles}, M., {Terlevich}, R., \& {Garcia-Pelayo}, J.-M. 1987,
  \mnras, 226, 849

\bibitem[{{Monreal-Ibero} {et~al.}(2007){Monreal-Ibero}, {Colina}, {Arribas},
  \& {Garc{\'{\i}}a-Mar{\'{\i}}n}}]{2007A&A...472..421M}
{Monreal-Ibero}, A., {Colina}, L., {Arribas}, S., \&
  {Garc{\'{\i}}a-Mar{\'{\i}}n}, M. 2007, \aap, 472, 421

\bibitem[{{Murray} {et~al.}(2010){Murray}, {Quataert}, \&
  {Thompson}}]{2010ApJ...709..191M}
{Murray}, N., {Quataert}, E., \& {Thompson}, T.~A. 2010, \apj, 709, 191

\bibitem[{{Noguchi}(1999)}]{1999ApJ...514...77N}
{Noguchi}, M. 1999, \apj, 514, 77

\bibitem[{{Osterbrock}(1989)}]{1989agna.book.....O}
{Osterbrock}, D.~E. 1989, {Astrophysics of gaseous nebulae and active galactic
  nuclei}, ed. {Osterbrock, D.~E.}

\bibitem[{{Pettini} \& {Pagel}(2004)}]{2004MNRAS.348L..59P}
{Pettini}, M., \& {Pagel}, B.~E.~J. 2004, \mnras, 348, L59

\bibitem[{{Pleuss} {et~al.}(2000){Pleuss}, {Heller}, \&
  {Fricke}}]{2000A&A...361..913P}
{Pleuss}, P.~O., {Heller}, C.~H., \& {Fricke}, K.~J. 2000, \aap, 361, 913

\bibitem[{{Rela{\~n}o} {et~al.}(2005){Rela{\~n}o}, {Beckman}, {Zurita},
  {Rozas}, \& {Giammanco}}]{2005A&A...431..235R}
{Rela{\~n}o}, M., {Beckman}, J.~E., {Zurita}, A., {Rozas}, M., \& {Giammanco},
  C. 2005, \aap, 431, 235

\bibitem[{{Rozas} {et~al.}(2006){Rozas}, {Richer}, {L{\'o}pez}, {Rela{\~n}o},
  \& {Beckman}}]{2006A&A...455..539R}
{Rozas}, M., {Richer}, M.~G., {L{\'o}pez}, J.~A., {Rela{\~n}o}, M., \&
  {Beckman}, J.~E. 2006, \aap, 455, 539

\bibitem[{{Samland} \& {Gerhard}(2003)}]{2003A&A...399..961S}
{Samland}, M., \& {Gerhard}, O.~E. 2003, \aap, 399, 961

\bibitem[{{Sandage} \& {Tammann}(1974)}]{1974ApJ...190..525S}
{Sandage}, A., \& {Tammann}, G.~A. 1974, \apj, 190, 525

\bibitem[{{Schmidt}(1959)}]{1959ApJ...129..243S}
{Schmidt}, M. 1959, \apj, 129, 243

\bibitem[{{Shapiro} {et~al.}(2009){Shapiro}, {Genzel}, {Quataert}, {F{\"o}rster
  Schreiber}, {Davies}, {Tacconi}, {Armus}, {Bouch{\'e}}, {Buschkamp},
  {Cimatti}, {Cresci}, {Daddi}, {Eisenhauer}, {Erb}, {Genel}, {Hicks}, {Lilly},
  {Lutz}, {Renzini}, {Shapley}, {Steidel}, \& {Sternberg}}]{Shapiro:2009sj}
{Shapiro}, K.~L., {et~al.} 2009, \apj, 701, 955

\bibitem[{{Smith} \& {Weedman}(1970)}]{1970ApJ...161...33S}
{Smith}, M.~G., \& {Weedman}, D.~W. 1970, \apj, 161, 33

\bibitem[{{Solomon} {et~al.}(1987){Solomon}, {Rivolo}, {Barrett}, \&
  {Yahil}}]{1987ApJ...319..730S}
{Solomon}, P.~M., {Rivolo}, A.~R., {Barrett}, J., \& {Yahil}, A. 1987, \apj,
  319, 730

\bibitem[{{Spitzer}(1942)}]{1942ApJ....95..329S}
{Spitzer}, Jr., L. 1942, \apj, 95, 329

\bibitem[{{Stone} {et~al.}(1998){Stone}, {Ostriker}, \&
  {Gammie}}]{1998ApJ...508L..99S}
{Stone}, J.~M., {Ostriker}, E.~C., \& {Gammie}, C.~F. 1998, \apjl, 508, L99

\bibitem[{{Swinbank} {et~al.}(2010){Swinbank}, {Smail}, {Longmore}, {Harris},
  {Baker}, {De Breuck}, {Richard}, {Edge}, {Ivison}, {Blundell}, {Coppin},
  {Cox}, {Gurwell}, {Hainline}, {Krips}, {Lundgren}, {Neri}, {Siana},
  {Siringo}, {Stark}, {Wilner}, \& {Younger}}]{2010Natur.464..733S}
{Swinbank}, A.~M., {et~al.} 2010, \nat, 464, 733

\bibitem[{{Swinbank} {et~al.}(2009){Swinbank}, {Webb}, {Richard}, {Bower},
  {Ellis}, {Illingworth}, {Jones}, {Kriek}, {Smail}, {Stark}, \& {van
  Dokkum}}]{Swinbank:2009tw}
---. 2009, \mnras, 400, 1121

\bibitem[{{Tacconi} {et~al.}(2010){Tacconi}, {Genzel}, {Neri}, {Cox}, {Cooper},
  {Shapiro}, {Bolatto}, {Bouch{\'e}}, {Bournaud}, {Burkert}, {Combes},
  {Comerford}, {Davis}, {Schreiber}, {Garcia-Burillo}, {Gracia-Carpio}, {Lutz},
  {Naab}, {Omont}, {Shapley}, {Sternberg}, \& {Weiner}}]{2010Natur.463..781T}
{Tacconi}, L.~J., {et~al.} 2010, \nat, 463, 781

\bibitem[{{Tenorio-Tagle} {et~al.}(1993){Tenorio-Tagle}, {Munoz-Tunon}, \&
  {Cox}}]{1993ApJ...418..767T}
{Tenorio-Tagle}, G., {Munoz-Tunon}, C., \& {Cox}, D.~P. 1993, \apj, 418, 767

\bibitem[{{Terlevich} \& {Melnick}(1981)}]{1981MNRAS.195..839T}
{Terlevich}, R., \& {Melnick}, J. 1981, \mnras, 195, 839

\bibitem[{Tiran {et~al.}(2011)Tiran, Lehnert, van Driel, Nesvadba, \&
  Matteo}]{Tiran:2011fk}
Tiran, L.~L., Lehnert, M.~D., van Driel, W., Nesvadba, N. P.~H., \& Matteo,
  P.~D. 2011

\bibitem[{{Toomre}(1964)}]{1964ApJ...139.1217T}
{Toomre}, A. 1964, \apj, 139, 1217

\bibitem[{{Tully} {et~al.}(2009){Tully}, {Rizzi}, {Shaya}, {Courtois},
  {Makarov}, \& {Jacobs}}]{2009AJ....138..323T}
{Tully}, R.~B., {Rizzi}, L., {Shaya}, E.~J., {Courtois}, H.~M., {Makarov},
  D.~I., \& {Jacobs}, B.~A. 2009, \aj, 138, 323

\bibitem[{{van Dam} {et~al.}(2006){van Dam}, {Bouchez}, {Le Mignant},
  {Johansson}, {Wizinowich}, {Campbell}, {Chin}, {Hartman}, {Lafon}, {Stomski},
  \& {Summers}}]{2006PASP..118..310V}
{van Dam}, M.~A., {et~al.} 2006, \pasp, 118, 310

\bibitem[{{van den Bergh} {et~al.}(1996){van den Bergh}, {Abraham}, {Ellis},
  {Tanvir}, {Santiago}, \& {Glazebrook}}]{van-den-bergh:1996:08}
{van den Bergh}, S., {Abraham}, R.~G., {Ellis}, R.~S., {Tanvir}, N.~R.,
  {Santiago}, B.~X., \& {Glazebrook}, K.~G. 1996, \aj, 112, 359

\bibitem[{{van Starkenburg} {et~al.}(2008){van Starkenburg}, {van der Werf},
  {Franx}, {Labb{\'e}}, {Rudnick}, \& {Wuyts}}]{2008A&A...488...99V}
{van Starkenburg}, L., {van der Werf}, P.~P., {Franx}, M., {Labb{\'e}}, I.,
  {Rudnick}, G., \& {Wuyts}, S. 2008, \aap, 488, 99

\bibitem[{{Wada} {et~al.}(2002){Wada}, {Meurer}, \&
  {Norman}}]{2002ApJ...577..197W}
{Wada}, K., {Meurer}, G., \& {Norman}, C.~A. 2002, \apj, 577, 197

\bibitem[{{Wisnioski} {et~al.}(2011){Wisnioski}, {Glazebrook}, {Blake},
  {Wyder}, {Martin}, {Poole}, {Sharp}, {Couch}, {Kacprzak}, {Brough},
  {Colless}, {Contreras}, {Croom}, {Croton}, {Davis}, {Drinkwater}, {Forster},
  {Gilbank}, {Gladders}, {Jelliffe}, {Jurek}, {Li}, {Madore}, {Pimbblet},
  {Pracy}, {Woods}, \& {Yee}}]{2011MNRAS.417.2601W}
{Wisnioski}, E., {et~al.} 2011, \mnras, 417, 2601

\bibitem[{{Wizinowich} {et~al.}(2006){Wizinowich}, {Le Mignant}, {Bouchez},
  {Campbell}, {Chin}, {Contos}, {van Dam}, {Hartman}, {Johansson}, {Lafon},
  {Lewis}, {Stomski}, {Summers}, {Brown}, {Danforth}, {Max}, \&
  {Pennington}}]{2006PASP..118..297W}
{Wizinowich}, P.~L., {et~al.} 2006, \pasp, 118, 297

\end{thebibliography}

\clearpage

\appendix

\label{app}
\section{Corrections to Clump \& \HII Region Measurements}
This appendix summarises the methods employed to calculate sizes, luminosities, and velocity dispersions from the literature, used for comparison with WiggleZ clumps. Details of the corrections we introduced to improve the accuracy of comparison are included.\\
\subsection{\HII Regions}

\textbf{SINGS measurements}\\
From the SINGS survey we measure \HII region size and luminosities from {NGC 24, NGC 628, NGC 925, NGC 1566, NGC 4254, NGC 3938, NGC 7552} as detailed in Section 3. These galaxies were chosen for the availability of high quality \halpha images and high numbers of \HII regions.

\subsection{Giant \HII regions}
\textbf{Gallagher \& Hunter 1983}\\
The \halpha flux is measured within a fixed 200 pc radius for each \HII region. This essentially yields the flux within the core radius. The authors estimate that there is up to a factor of two error in the flux measurements. Velocity dispersions are corrected for Doppler broadening and instrumental broadening. We recalculate the luminosity and diameters with $H_{0}=70$ km s$^{-1}$ Mpc$^{-1}$ from $H_{0}=50$  km s$^{-1}$ Mpc$^{-1}$ originally used in the paper, using new distances from the Extragalactic Distance Database. \\

\hspace*{-0.65cm}\textbf{Arsenault 1988}\\
Recalculation of \HII region luminosities and diameters with $H_{0}=70$ km s$^{-1}$ Mpc$^{-1}$ and new distances from the Extragalactic Distance Database is readily possible as the authors provide the original measurements of angular diameter and \halpha flux. Velocity dispersions are not measured for this sample. \\

\hspace*{-0.65cm}\textbf{Bastian et al. 2006}\\
Six \HII regions are observed in the Antennae galaxies, NGC 4038 and NGC 4039, with the VLT-VIMOS IFS. The H$\gamma$, H$\beta$, and \OIII~emission lines are observed. The Antennae is classified as a LIRG. Luminosities are inferred from \hbeta and converted to \halpha assuming case B recombination (10 000 K).
We recalculate the luminosity and diameters with $H_{0}=70$ km s$^{-1}$ Mpc$^{-1}$ from $H_{0}=75$  km s$^{-1}$ Mpc$^{-1}$ originally used in the paper, with a new distance of $D=19.2$ Mpc from the Extragalactic Distance Database.\\

\hspace*{-0.65cm}\textbf{Rozas et al. 2006}\\
This sample includes only the intrinsically brightest \HII regions within NGC 157, NGC 925, NGC 3631, NGC 6764, NGC 3344, NGC 4321, NGC 5364, NGC 5055, NGC 5985, and NGC 7479. The purpose of the selection was to choose the youngest, most massive normal star clusters to minimise the processes available to drive the \HII region away from virial equilibrium. To find an accurate velocity dispersion the authors determine the optimal number of Gaussian components to fit to each line profile using the component which contributes the most luminosity in each \HII region. For region size they adopt one half the total radius given in \HII region catalogues as specified by \cite{2005A&A...431..235R}, which is based on the 40\% isophote method of McCall 1990, which is equivalent to our core method. The velocity dispersion is measured within this radius and was corrected for natural broadening, Doppler broadening, and instrumental broadening.
We recalculate the luminosity and diameters with $H_{0}=70$ km s$^{-1}$ Mpc$^{-1}$ from $H_{0}=75$  km s$^{-1}$ Mpc$^{-1}$ originally used in the paper, with new distances from the Extragalactic Distance Database. \\

\hspace*{-0.65cm}\textbf{Monreal-Ibero et al. 2007}\\
The regions selected from this paper are tidal dwarf galaxy candidates. However, they are selected in a similar manner to other high-redshift and low-redshift regions used for comparison in this paper and as such we include them as giant \HII regions. They are selected within 5 local ULIRGs with WFPC2 and IFS data. Isophotal and effective radii are measured for all regions. We use the effective radii as it better approximates the core method. Luminosity is recalculated from the given flux values without extinction correction.

\subsection{High-Redshift Clumps}
\textbf{SINS Survey: Genzel et al. (2011) \& F\"{o}rster-Schrieber et al. (2011)}\\
The clump data taken from \cite{2011ApJ...733..101G} are for BX482-clumpA, ZC782941-clumpA, ZC406690-clumpA, ZC406690-clumpB, and ZC406690-clumpC. Integrated clump properties such as BX599-all and D3a15504-clumpsA-F are excluded. Clump radii are given by the HWHM from Gaussian fits to clumps in velocity channels of the IFS data cube. The instrumental resolution is subtracted in quadrature. We convert the HWHM to the r.m.s. of a Gaussian profile, the radius used for the core method, by $r_{\mathrm{core}} = \frac{2 \mathrm{HWHM}}{\sqrt{2\log{2}}}$. The velocity dispersion of the clumps is the intrinsic local velocity dispersion after removal of beam-smeared rotation and instrumental resolution. 

The clump data taken from \cite{2011ApJ...739...45F} are exclusively from  BX482. Clump-1 in this sample is defined to be the same clump as BX482-clumpA in \cite{2011ApJ...733..101G}. In \cite{2011ApJ...739...45F} the flux, $F_\mathrm{obs}=3.92\pm0.05 \times 10^{-17}$ ergs s$^{-1}$ cm$^{-2}$ is measured within $r_{\mathrm{phot}}=$ 2.3 kpc and has $R_{\mathrm{HWHM}}=0.96$ kpc in \cite{2011ApJ...733..101G} the $R_{\mathrm{HWHM}}=1$ kpc and $F_\mathrm{obs}=3.5\times 10^{-17}$ ergs s$^{-1}$ cm$^{-2}$. For comparison purposes we use $r_\mathrm{phot}$ radii.\\

\hspace*{-0.65cm}\textbf{Swinbank et al. 2009 \& Jones et al. 2010}\\
\cite{Swinbank:2009tw} and \cite{Jones:2010uf} publish clumps in $L_{*}$ lensed galaxies at $z\sim2-4$ from IFS data. They derive sizes using the isophote method.  We correct the cosmology to $H_0=70$ km s$^{-1}$ Mpc$^{-1}$ and $\Omega_{\mathrm{M}}=0.27$, $\Omega_{\Lambda}=0.73$. 

\clearpage
\label{lastpage}

\end{document}